\documentclass[twocolumn]{aa} 
\usepackage{graphicx}
\usepackage{amsmath}
\usepackage[]{natbib}
\usepackage{txfonts}
\begin{document}

   \title{Massive star evolution in close binaries:}

\subtitle{conditions for homogeneous chemical evolution}

   \author{H.F. Song,
          \inst{1,3}
          G. Meynet\inst{2*},          
           A. Maeder\inst{2},
          S. Ekstr\"om\inst{2} 
          \and
          P. Eggenberger\inst{2}}
\authorrunning{Song et al.}
\institute{College of Science, Guizhou University, Guiyang, Guizhou Province, 550025, P.R. China
\and
Geneva Observatory, Geneva University, CH-1290 Sauverny, Switzerland
\and
Key Laboratory for the Structure and Evolution of Celestial Objects, Chinese Academy of Sciences, Kunming 650011             
$^{*}$Corresponding author,  \email{georges.meynet@unige.ch}
}

   \date{Received; accepted }


  \abstract
   {}
   {
   We investigate the impact of tidal interactions, before any mass transfer, on various properties of the stellar models. 
    We study the conditions for obtaining homogeneous evolution
   triggered by tidal interactions, and for avoiding any Roche lobe overflow during the
   Main-Sequence phase. By homogeneous evolution, we mean  stars evolving with a nearly uniform chemical composition from the center to the surface.
   }
  { We consider the case of  rotating stars computed with a strong coupling mediated by an interior magnetic field. Models with initial masses between 15 and 60 M$_\odot$, for metallicities between 0.002 and 0.014, with initial rotation equal to 30\% and 66\% the critical rotation on the ZAMS are computed for single stars and for stars in close binary systems. Close binary systems with initial orbital periods equal to 1.4, 1.6 and 1.8 days and a mass ratio equal to 3/2 are considered.
}
 {
 In models without any tidal interaction (single stars and wide binaries), homogeneous evolution in solid body rotating models is obtained when two conditions are realized: the initial rotation must be high enough, the loss
 of angular momentum by stellar winds should be modest. This last point favors metal-poor fast rotating stars. In models with tidal interactions, homogeneous evolution
 is obtained when rotation imposed by synchronization is high enough (typically a time-averaged surface velocities during the Main-Sequence phase above 250 km s$^{-1}$), whatever the mass losses. 
 We give plots indicating for which masses of the primary and for which initial periods,  the conditions for the homogenous evolution and for the avoidance of the Roche lobe overflow are met, this for different initial metallicities and rotations. In close binaries, mixing is stronger at higher than at lower metallicities. Homogeneous evolution is thus favored at higher metallicities. Roche lobe overflow avoidance is favored at lower metallicities due to the fact that stars with less metals remain more compact. 
 We study also the impact of different processes for the angular momentum transport on the surface abundances and velocities in single and close binaries.
In models where strong internal coupling is assumed, strong surface enrichments are always associated to high surface velocities in binary or single star models. In contrast, models
 computed with mild coupling may produce strong surface enrichments associated to  low surface velocities. This observable difference can be used to probe different models for
 the transport of the angular momentum in stars. Homogeneous  evolution is more easily obtained in models (with or without tidal interactions) with solid body rotation. 
 }
 {Close binary models may be of interest for explaining homogeneous massive stars, fast rotating Wolf-Rayet stars, and
 progenitors of long soft gamma ray bursts, even at high metallicities.
 }


   \maketitle
%

\section{Introduction}

A significant fraction of massive stars may belong to close binary systems \citep[see e.g.][]{Sana12, Sana13}. Tidal interaction in close binary systems can
change the evolution of the primary well before any mass transfer \citep{deMink09a, Song13}. In particular, strong mixing can be induced leading
to a nearly homogeneous evolution.
Homogeneous evolution of stars is a topic of interest for many reasons: 
models and observations suggest the existence of such stars \citep{Maeder87, Martins09, Martins13}. Second, when occurring in close binary systems, such a homogeneous evolution may imply that the Roche lobe overflow will be avoided
\citep{deMink09a}. Third, this type of evolution may be of interest as a possible scenario leading to long soft gamma-ray burst
\citep{Yoon06}. Fourth, these stars are powerful sources of ionizing photons \citep{Meynet08, Levesque12, Leitherer14} and if frequent enough in the early Universe could have
contributed to its reionization.
Finally, such an evolution may be also of interest in the frame of the studies aiming at understanding the anticorrelations
between the abundances of some light elements like oxygen and sodium observed at the surface of a large fraction of stars in globular clusters \citep[see discussions of the observational evidences and proposed models in the reviews by][]{Gratton04, Gratton12}.

Homogeneous evolution can be triggered by various mechanisms:
\begin{itemize}
\item 1) Very massive stars have so large convective cores that their evolution can be considered as nearly homogeneous \citep{Maeder80, Yusof13}
\item 2) Strong internal mixing in radiative zones produce homogeneous evolution \citep{Maeder87}. 
\end{itemize}
In this work, we shall focus on models with strong mixing in the radiative zones. This strong mixing can be due either to fast rotation and/or due to some Tidally Induced Shear Mixing (TISM, see more below). 

Fast rotation can be the result of initial conditions, the star rotating fast on the ZAMS, or due to some accelerating mechanism
induced by tidal forces, mass accretion or merging of stars. 

TISM  does not require any high initial rotation, it needs only the build up of strong shear gradients in the star. It will occur only in close binary systems, either when the star is spin-up or spin-down \citep{Song13}.
It can occur only in stars which have mild internal transport processes of angular momentum (like for instance transports by shear and meridional currents). 

In case
the transport of angular momentum is quite fast \citep[like the one triggered by a strong internal magnetic field, as in the work of][and as in most of the models computed in the present work]{deMink09a}, at every time, the star rotates nearly as a solid body and therefore the transport by the shear becomes negligible! The tidal interaction in that case mainly serves as a process to give the star a fast enough rotation to drive a homogeneous evolution. In the present work, we focus on this type of models. As such, it is a follow up of the study made by \citet{deMink09a}, with however
the following differences:
\begin{itemize}
\item In \citet{deMink09a}, the space of parameters studied are massive binaries with initial masses 20 + 15 M$_\odot$ and 50 + 25 M$_\odot$,
with a composition representative of the Small Magellanic Cloud (mass fraction of heavy elements $Z$=0.0021) and with initial orbital periods between 1.1 and 4 days. Here we study binaries with initial masses 60 + 40 M$_\odot$, 50 + 33.3 M$_\odot$, 40 + 26.7 M$_\odot$,
30 + 20 M$_\odot$, and 15 + 10 M$_\odot$ for initial metallicities Z=0.002, 0.007 and 0.014 and initial orbital periods equal to 1.4, 1.6 and 1.8 days.
\item In \citet{deMink09a}, the synchronization phase was not explicitly computed. Based on an analytical estimate of the synchronization timescale, these authors showed that this time  is very short, less than 1\% of the Main-Sequence lifetime for the parameters considered in their work. Therefore, they began their computation 
assuming that the system is synchronized, the stars having rotation rates such that their spin period is equal to the orbital period.
Here we computed explicitly the synchronization phase accounting for the various transport processes. We computed models with high initial surface equatorial velocities between 360 and 420 km s$^{-1}$ on the ZAMS (see Tables~A.1 and A.2), that correspond to the cases of spin-down by the tidal forces and models with moderate initial
surface equatorial velocities between 150 and 175 km s$^{-1}$ on the ZAMS, that correspond to the cases of spin-up by the tidal forces.
\item The physical ingredients of our models are different from those used in \citet{deMink09a} (see more in Section 2 below). 
Also, in the Appendix B, we explore the consequences of tidal interactions when two different
transport mechanisms for the angular momentum are considered. One which produces a mild coupling between the core and the envelope mediated by only shear and meridional currents and the other
one, which is a strong coupling, mediated by an internal magnetic field. As we will see the consequences of the tidal interactions are very different. 
\end{itemize} 
In Section 2, we discuss the physical ingredients of the models. 
The results for 40 M$_\odot$ models, single and in close binaries, are presented in Section 3.
The models for stars with masses between 30 and 60 M$_\odot$, with and without tidal interaction are discussed in Section 4.
Section 5 discusses the conditions for obtaining a homogeneous evolution and for avoiding Roche lobe overflow in a close binary system.
Section 6 synthesizes the conclusions and indicates some further works both observational and theoretical that appear interesting to be done along this line of research.
The Appendix A present tables indicating some properties of the stellar models computed in this work.
The Appendix B is devoted to a discussion of the effects of tidal interactions on systems composed of a 15 and 10 M$_\odot$ for two different treatments of the transport of the angular momentum. 

\section{Physical ingredients of the models}

Except for what concerns the effects of rotation and of the tidal forces, the present models use the same physical ingredients as in the works by \citet{Ekstrom12, Georgy13a, Georgy13b}. In particular, we use the Schwarzshild criterion with a moderate overshoot for computing the size of the convective core. The overshooting parameter is
taken equal to 10\% the pressure scale height estimated at the border of the convective core given by the Schwarzschild's criterion. Note that
the models by  \citet{deMink09a} use a much higher overshooting parameter of 35.5\% the pressure scale height estimated at the border of the Ledoux's boundary, based on calibrations by \citet{Brott2011}.

The impact of the tides in close binary system is accounted for as in \citet{Song13}.
A significant difference however with respect to the work by \citet{Song13} is that
here we assume a strong internal coupling mediated by an internal magnetic field. More precisely we performed the computation including the equations of the Tayler-Spruit dynamo  \citep{Spruit02} as given in \citet{MM05}.
In this framework, for the mixing of the chemical species, we use the following equations:
$$\rho \frac{\text{d}X_i}{\text{d}t} = \frac{1}{r^2} \frac{\partial}{\partial r} \left( \rho r^2 \left[ D + D_\text{eff} \right] \frac{\partial X_i}{\partial r} \right) + \left( \frac{\text{d}X_i}{\text{d}t}\right)_\text{nucl},$$
with $\rho$ the density, $X_i$, the mass fraction of element $i$, $t$, the time, $r$, the radius, $D$ and $D_\text{eff}$, the diffusion coefficients (see below). The last last term on the right hand side expresses the changes of the mass fraction of element $i$ resulting from nuclear reactions.
In the above expression, $D=D_\text{shear} + D_\text{magn}$ with
$$D_\text{shear}=f_\text{energ} \frac{H_P}{g\delta}\frac{K}{\left[\frac{\varphi}{\delta}\nabla_\mu + \left( \nabla_\text{ad} - \nabla_\text{rad} \right)\right]} \left( \frac{9\pi}{32}\ \Omega\ \frac{\text{d} \ln \Omega}{\text{d} \ln r} \right)^2$$
where $K = \frac{4ac}{3\kappa}\frac{T^4\nabla_\text{ad}}{ \rho P \delta}$, and with  $f_\text{energ} = 1$, and $\varphi = \left( \frac{\partial \ln\rho}{\partial \ln\mu} \right)_{P,T} = 1$. The quantity $H_P$ is the pressure scale height, $g$, the gravity, $\delta=-({\partial \ln T / \partial \ln \rho})_{\mu, P}$, $\nabla_\text{ad}$, $\nabla_\text{rad}$ and $\nabla_\mu$ are respectively the adiabatic, radiative temperature gradients and the mean molecular weight gradient,  $\Omega$, the angular velocity, $T$, the temperature, $P$, the pressure, $\kappa$, the opacity, $a$, the radiation constant and $c$, the velocity of light. We have also
$$D_\text{eff}= \frac{1}{30} \frac{\left| r\ U(r) \right|^2}{D_\text{h}}$$
with
$$D_\text{h}= \frac{1}{c_\text{h}}\ r\ \left| 2\,V(r) - \alpha\,U(r) \right|$$
where $\alpha = \frac{1}{2} \frac{\text{d} \ln (r^2 \bar{\Omega})}{\text{d} \ln r}$ and $c_\text{h} = 1$,  $U(r)$ and $V(r)$ are the radial dependence of respectively the vertical and horizontal  
component of the meridional circulation velocity.
To define the magnetic diffusivity $D_\text{magn}$, we first have to obtain the Alfven frequency $\omega_\text{A}$ by solving the equation
$$\frac{r^2\Omega}{\left(\frac{\text{d}\ln\Omega}{\text{d}\ln r}\right)^2\,K}\,\left( N_\mu^2+N_T^2 \right)\,x^4 - \frac{r^2\,\Omega^3}{K}\,x^3 +2N_\mu^2\,x -2\Omega^2\left( \frac{\text{d}\ln\Omega}{\text{d}\ln r} \right)^2 = 0$$
with $x=\left(\frac{\omega_\text{A}}{\Omega}\right)^2$, $N^2_T={g\delta \over H_P} (\nabla_{\rm ad}-\nabla_{\rm rad})$ and $N^2_\mu={{g\varphi \over H_P}}\nabla_\mu$.
$D_\text{magn}$ comes then as:
$$D_\text{magn} = \frac{r^2\,\Omega}{\left( \frac{\text{d}\ln\Omega}{\text{d}\ln r} \right)^2}\,\left(\frac{\omega_\text{A}}{\Omega}\right)^6.$$

For the transport of the angular momentum, we use the following equation
$$\rho \frac{\text{d}}{\text{d}t} \left( r^2 \Omega \right)_{M_r} =
   \frac{1}{r^2} \frac{\partial}{\partial r} \left( \rho r^4 D \frac{\partial \Omega}{\partial r} \right)
$$
with $D=D_\text{shear} + D_\text{magn,$\Omega$} + D_\text{circ,H}$. In this expression,
$$D_\text{circ,H} = \left| \frac{r\,U(r)}{5\ \frac{\text{d} \ln \Omega}{\text{d} \ln r}} \right|$$
and
$$D_{\text{magn,}\Omega} = \frac{r^2\,\Omega}{\left| \frac{\text{d}\ln\Omega}{\text{d}\ln r} \right|}\,\left( \frac{\omega_\text{A}}{\Omega} \right)^3\,\frac{\Omega}{N_\text{Vais}}$$
with the general Brunt-V\"ais\"al\"a frequency
$$N_\text{Vais}^2 = \frac{D_\text{magn}}{2\,K}\,\frac{g\,\delta}{H_P}\,\left|\nabla_\text{rad}-\nabla_\text{ad}\right| + \frac{g\,\nabla_\mu}{H_P}.$$
We consider that the minimum shear for the dynamo to work is given by
$$\frac{\text{d}\ln\Omega}{\text{d}\ln r}>q_\text{min}$$
where $q_\text{min}=\left(\frac{N_\text{Vais}}{\Omega}\right)^{(7/4)}\,\left(\frac{D_\text{magn} }{r^2\,N_\text{Vais}}\right)^{(1/4)}$.


As can be seen from the expression for the transport of the angular momentum, 
we accounted for the effect of meridional currents on the transport of the angular momentum through a diffusive equation instead of an advective one as we usually do in models without magnetic field.  We can justify this way of doing by the following arguments: first, in models with magnetic field, the distribution of angular momentum inside the star
is dominated by the processes linked to the magnetic field which is a diffusive one and no long by the meridional currents (which is an advective process); Second, the transport by a magnetic field is so efficient that it imposes at every instant a nearly flat distribution of the angular velocity inside the star. Actually the present models are equivalent to solid-body rotating models. Would we have accounted for the meridional currents as an advective process, this would not have changed
that conclusion and we checked numerically that point. Indeed, 
in \citet{MM05}, we checked that nearly solid-body rotation is achieved in models where the meridional currents are treated as an advective process and where
the major processes governing the transport of the angular momentum are those associated to the magnetic instabilities.
We can note that keeping the advection by meridional currents for the transport of the angular momentum imposes the use of extremely short time steps and thus is very costly in CPU time. This would have made impossible
the computation of the $\sim$100 different stellar models that we discuss in the present work.

 Let us note that the dynamo theory proposed by \citet{Spruit02} has been
criticized by \citet{Zahn07}. These last authors consider that  the analytic treatment by \citet{Spruit02} is too simplified to be applied to the real stellar situation. On the other hand, 
the present models can be seen as a study of solid body rotating models, whatever the physical cause is held responsible for the building up of this solid body rotation.
Actually most of the chemical mixing is not due to the specific dynamo process but to the meridional currents expected to be present in solid
body rotating stars. Let us remind here that the concomitant effects of meridional currents and of the strong horizontal turbulence
allow to consider the mixing of the chemical species driven by these processes  through a diffusive equation (diffusion coefficient $D_{\rm eff}$). Thus for the chemical
mixing, in contrast to angular momentum, a diffusive treatment is anyway the one that has to be applied.


Computations  of the models indicated in Sect.~1 were performed
until the primary fills in its Roche lobe. In case, the primary does not succeed to fill in its Roche lobe, the computation was performed until the end of the
Main-Sequence phase. For purpose of comparison, computations of single star evolution have been performed for initial masses equal to the mass of the primaries and for the same initial rotational velocities as the one considered in the close binary systems.

As we shall see in the following, some of our models may enter a phase during which the mass fraction of hydrogen at the surface becomes lower than 0.3 and the effective temperature is larger than 4.0.
In that case, we consider our model star to be a WR$_e$, {\it i.e.} as a Wolf-Rayet star defined from stellar evolution criteria. The WR$_e$ may not be exactly equivalent to the
bona fide WR stars classified according to spectroscopic criteria \citep[see for instance the discussion in][]{Groh2014}, however there is some good overlap between these two definitions.

We have calculated grids of single and binary stars in the range of masses  30 to 60 M$_\odot$, with mass ratios equal to 3/2. We have  considered 3 metallicities Z=0.014, 0.007, 0.002  to cover objects in the solar neighborhood, the LMC and SMC. 
A variety of orbital periods and of initial axial rotation velocities is considered, so that on the whole the evolutions of 96 different systems are calculated.



\section{The 40 M$_\odot$ stellar models}

Before to present a global overview of all the results, let us first discuss the specific cases of 40 M$_\odot$ models.

\begin{figure*}
   \centering
    \includegraphics[width=8.9cm]{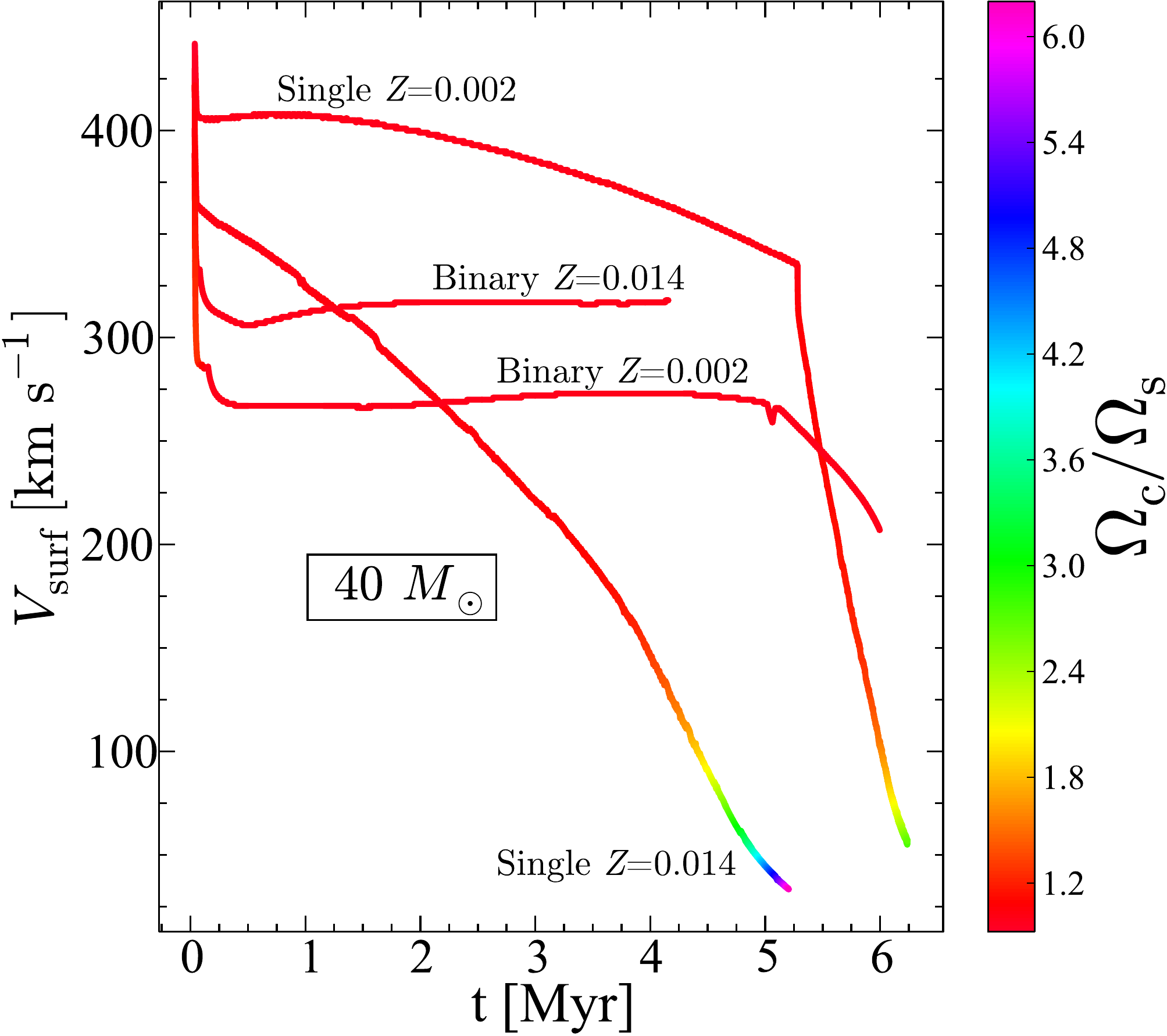}\hfill\hfill  \includegraphics[width=8.9cm]{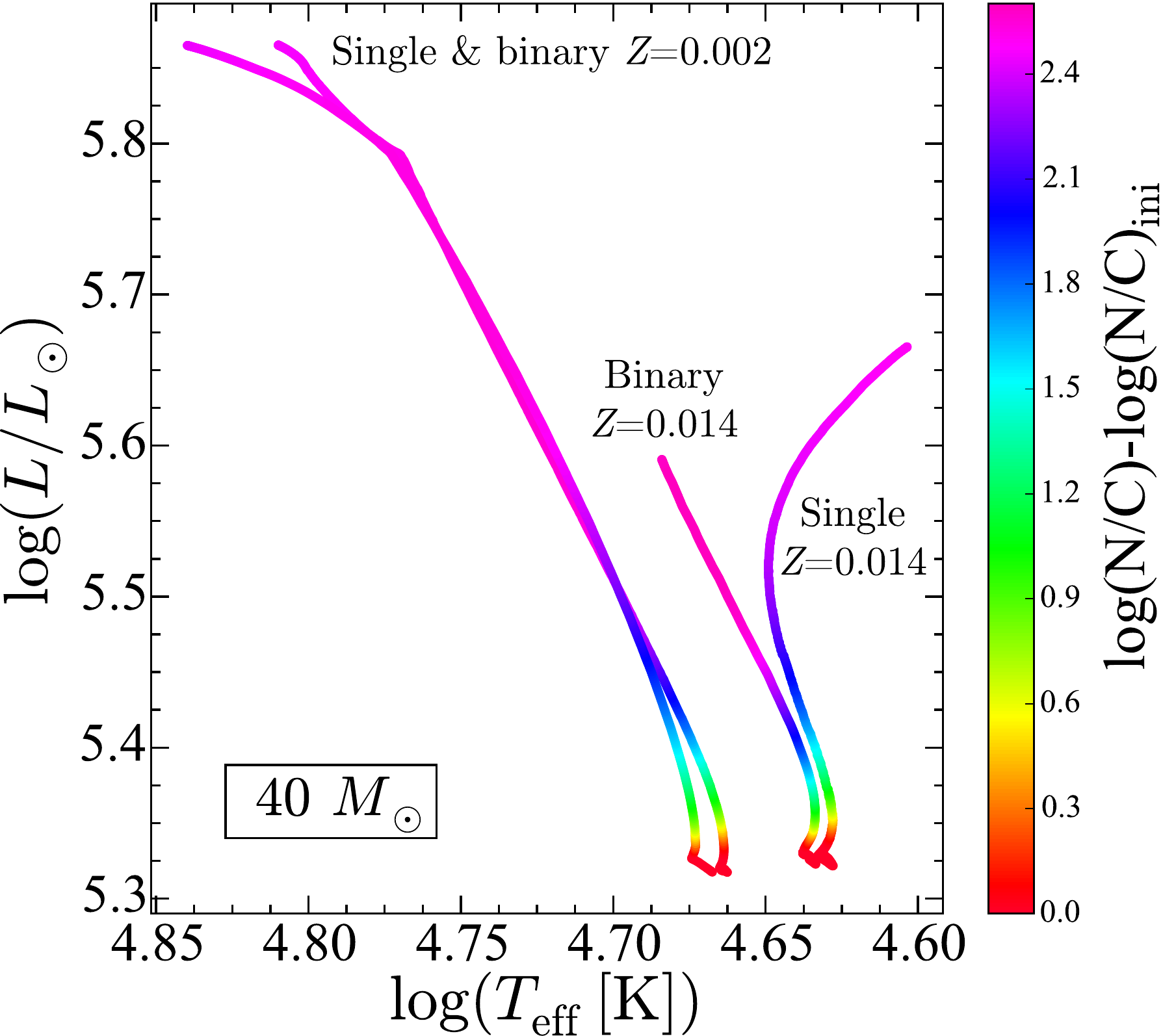}   
      \caption{{\it Left panel:}  Evolution of the surface equatorial velocity as a function of time for 40 M$_\odot$ models. The colors indicate the ratio between the central
      angular velocity and the surface one. The orbital period for the binary models is 1.4 days, the secondary has an initial mass equal to 26.7 M$_\odot$.
      {\it Right panel:} Evolutionary tracks in the HRD for the same models as in the left panel. 
      The colors indicate the N/C  ratio at the surface normalized to the initial one.
      In the binary systems, the stars have initially an angular velocity superior to the orbital one and thus the tidal forces will spin-down the star (see Table A.2).
}
         \label{m40sd}
   \end{figure*}

\begin{figure*}
   \centering
    \includegraphics[width=8.9cm]{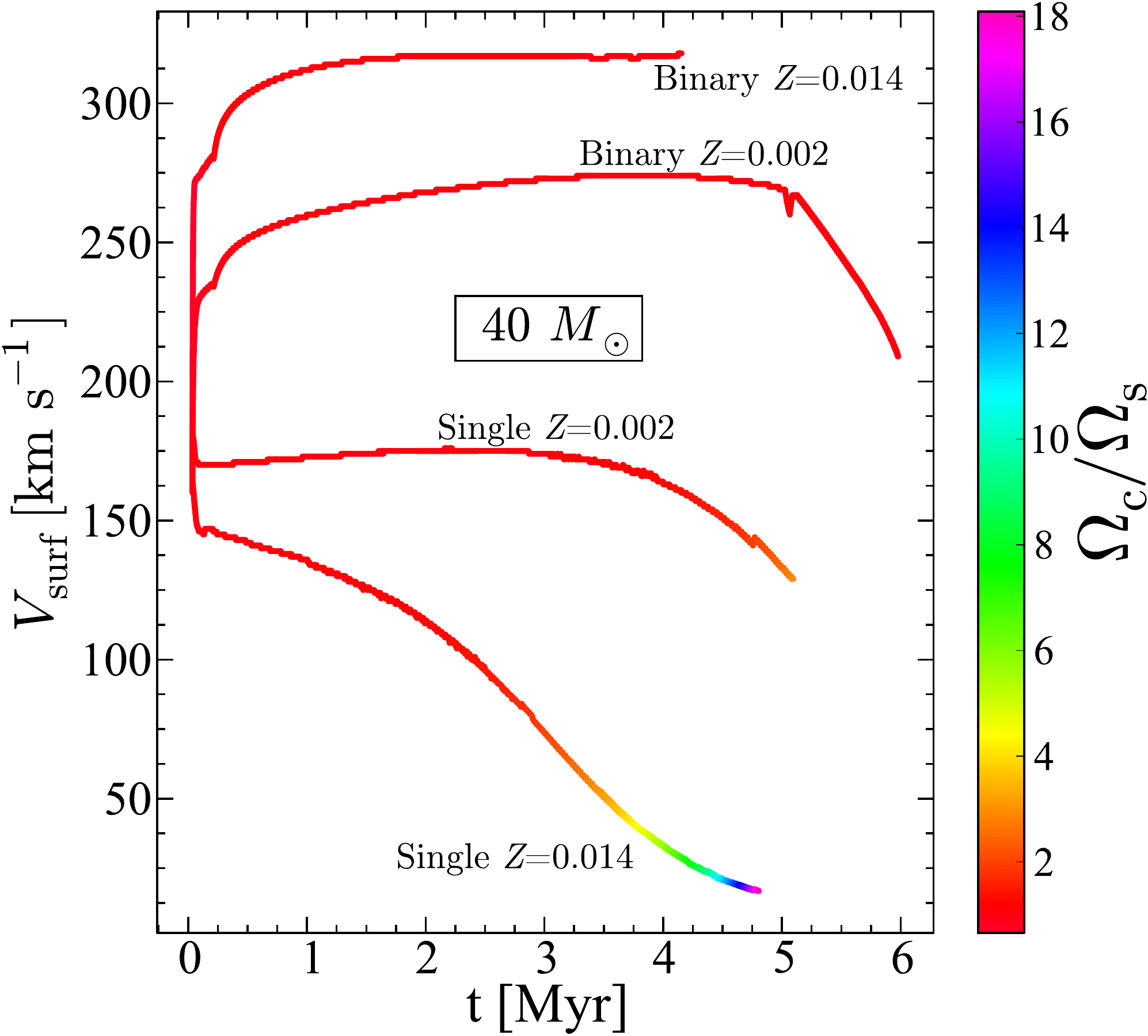}\hfill\hfill  \includegraphics[width=8.9cm]{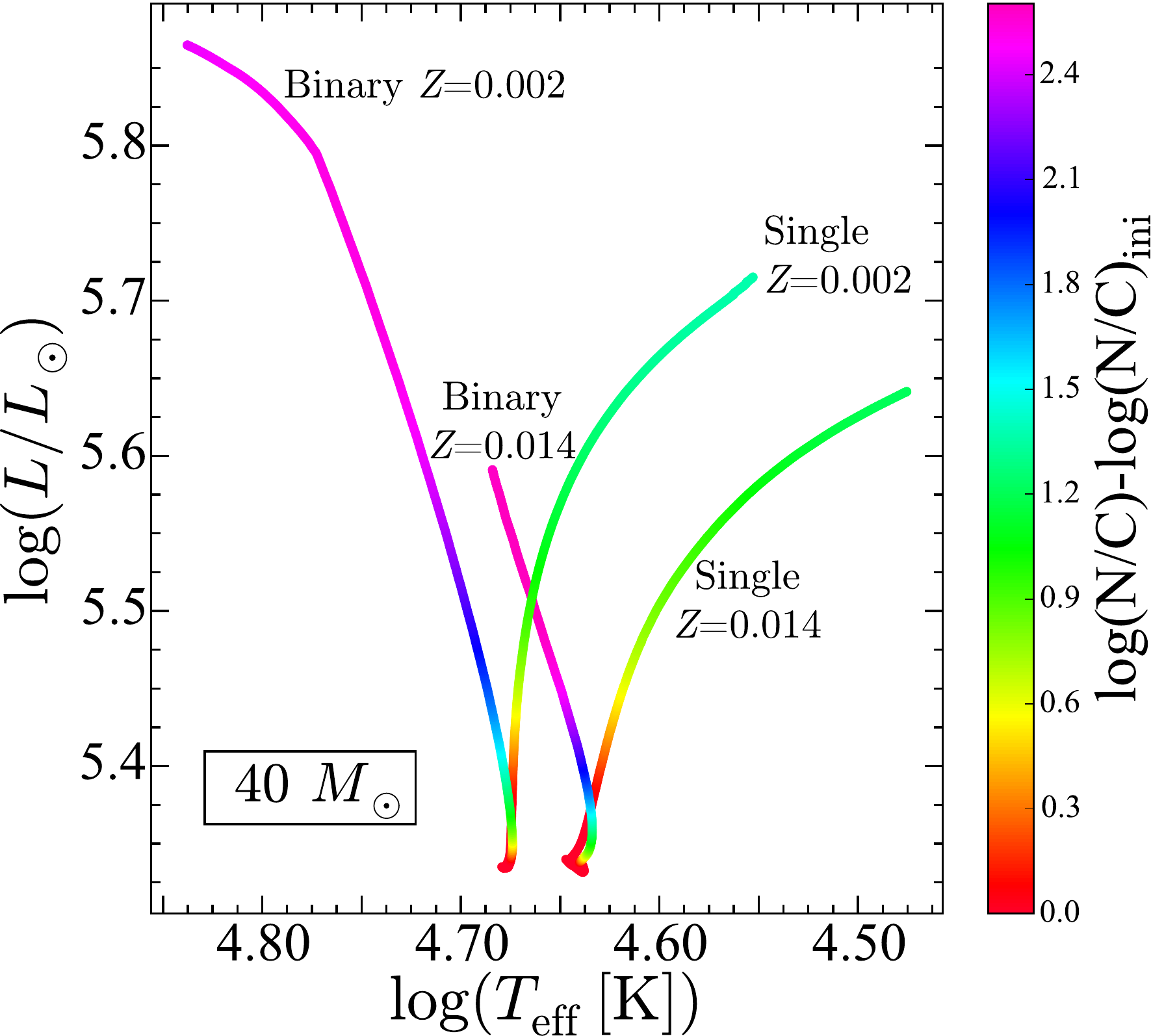}   
      \caption{Same as Fig.~\ref{m40sd} for stars starting with a lower initial rotation (see Table A.2). In binary systems, the primary will be spun-up by the tidal forces.}
         \label{m40su}
   \end{figure*}

In Figs.~\ref{m40sd} and \ref{m40su}, the evolutions of the surface equatorial velocities (left panel) and 
of the tracks in the theoretical HR diagram are shown (right panel) for both single and close binary 40 M$_\odot$ stellar models (orbital period is 1.4 days). In Fig.~\ref{m40sd}, in binary systems, the primary star begins with an angular velocity larger than the orbital one and thus we have  a spin-down case, while in Fig.~\ref{m40su}, we have a spin-up case. 

Let us first discuss the single star models.
The most striking effect to be noted is that at the metallicity $Z$=0.014, the surface velocities
strongly decrease as a function of time as a result of the angular momentum losses driven by the stellar winds. This occurs for initially fast
and moderately rotating models. The contrast with the angular velocity at the center and the angular velocity at the surface is quite small
illustrating the strong coupling induced by the magnetic field. The core begins to rotate significantly faster than the envelope only 
at the very end of the Main-Sequence. This comes from the ever increasing inhibiting effect of the $\mu$-gradient at the border of the convective core when evolution goes on. The $\mu$-gradients decrease the transport by magnetic fields and therefore the strong coupling. 
The contrast is stronger in models with Z=0.014 than in the model at Z=0.002. This comes from the fact that once the coupling
between the core and the envelope is no longer achieved, the stronger mass loss rates at higher metallicities increase faster the contrast
between the central and the surface rotation rate.

Looking at the tracks in the HR diagram, we see that the fast rotating model at high metallicity
first begins to follow the typical blueward evolution characterizing a model evolving homogeneously. Then, due to the spin down by the stellar
winds, the track bends to the red. The starting point of the model with $Z=0.002$ 
is shifted to the blue because the star is more compact. This is mainly an
opacity effect \citep[see e.g. the discussion in][]{Mowlavi98}. We see that since that model loses much less angular momentum, the mixing
driven by rotation remains quite high and the star follows a nearly homogeneous evolution.

In case of models starting with a lower initial rotation (see Fig.~\ref{m40su}), we have for the evolution of the surface velocities qualitatively similar behaviors as those obtained for initially fast rotating stars, the level of the surface rotation for single stars  being simply shifted to lower values. This is the reason why
for both metallicities considered in these plots, the evolutionary tracks for single stars evolve to the red part of the HRD as is normal for stars undergoing no
or weak mixing in the radiative envelope.

Let us now discuss the close binary models.
One of the main impact of the tidal interactions is to change the evolution of the surface rotation rates as can be seen in 
Figs.~\ref{m40sd} and \ref{m40su}. We see that the changes of the surface rotation due to tidal forces is very rapid both in case
of spin-down (Fig.~\ref{m40sd}) and in case of spin-up (Fig.~\ref{m40su}).

Once synchronization is achieved,
the surface velocities keep a more or less constant value for a significant part of the track (see Figs.~\ref{m40sd} and \ref{m40su}). 
This is of course a consequence of the fact that
the surface angular velocity is locked by the tidal interaction in a small domain around the orbital angular momentum velocity. 
We see also that since the orbital angular velocity considered is the same for the case of spin-up and spin-down, the surface velocities
obtained after synchronization are very similar.

The models of 40 M$_\odot$ with tidal interactions are more mixed than models without tidal interactions, both in case of spin-down and in case of spin-up for the orbital periods considered here. It is of course quite natural to expect that the spin-up model is more mixed than the
single star. It is a little less obvious in the case of spin-down. We can see that at $Z=0.014$
the spin down model has a surface rotation velocity which
is larger than the single star model. This results from the fact that the single star model loses much more angular momentum than
the model in the close binary. The single star loses angular momentum by stellar winds.
The star in the close binary also loses angular momentum by stellar winds, but the tidal interactions
compensate for these losses  in order to achieve synchronization.
This acceleration occurring in close binary is done through a transfer of angular momentum from the orbital reservoir to the star reservoir. Note that
the amount of angular momentum in stars is very small with respect to the orbital angular momentum, so that this transfer hardly changes the
orbital period and the distance between the components. 
So for the conditions considered here,  the spin-down model in a close binary system will be  
more mixed than the single star mass-losing model.

At $Z=0.002$, the surface velocity of the spin-down model is most of the time below the surface velocity of the single star. 
So here we should expect less mixing in the spin down close binary model.
However this model  follows a homogeneous evolution
as the faster rotating model with no tidal interaction. What happens here is that both models (with and without tidal interaction) keep
sufficient angular momentum to be homogeneously mixed.

\begin{figure*}
   \centering
    \includegraphics[width=8.9cm]{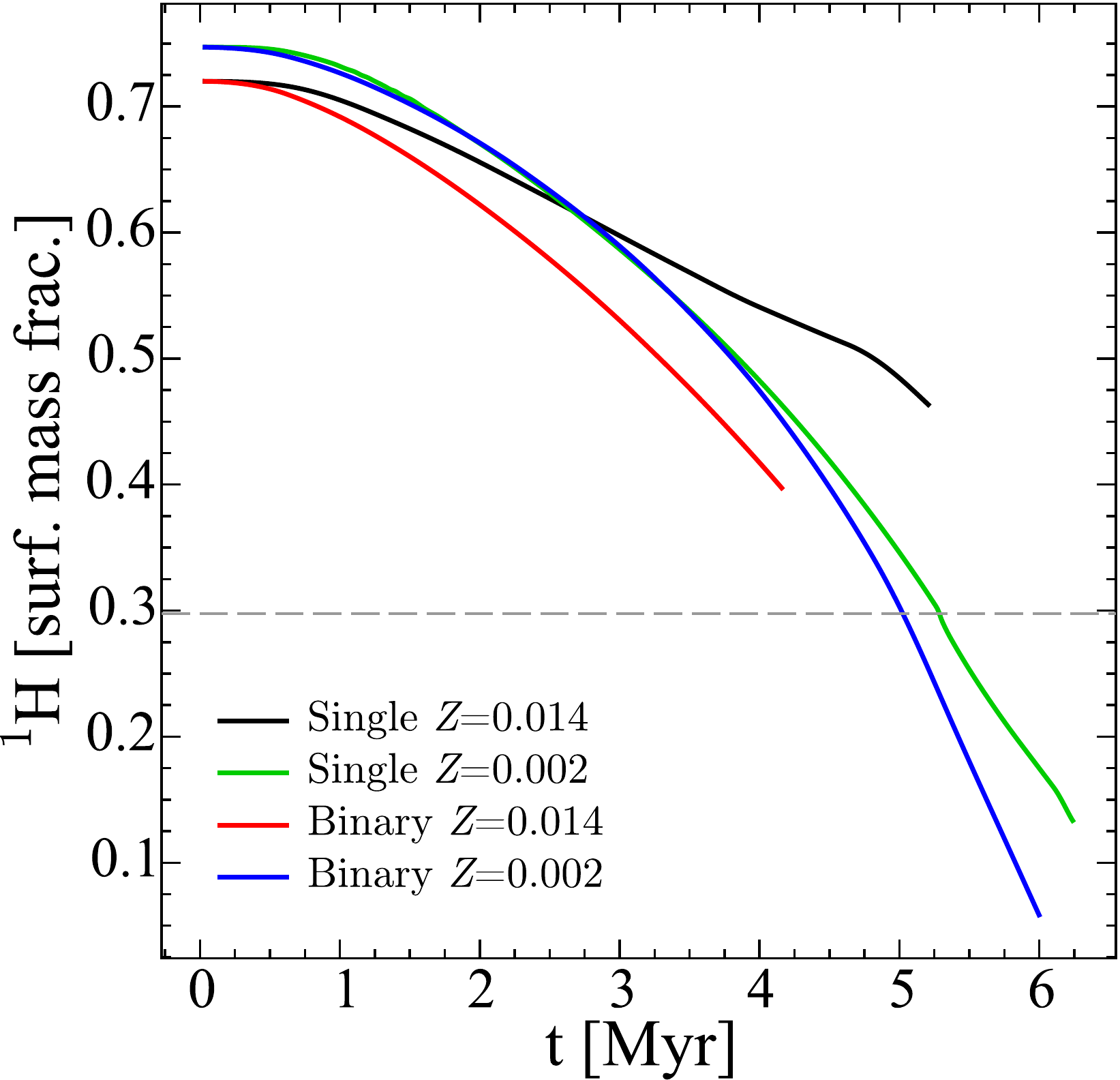}\hfill\hfill  \includegraphics[width=8.9cm]{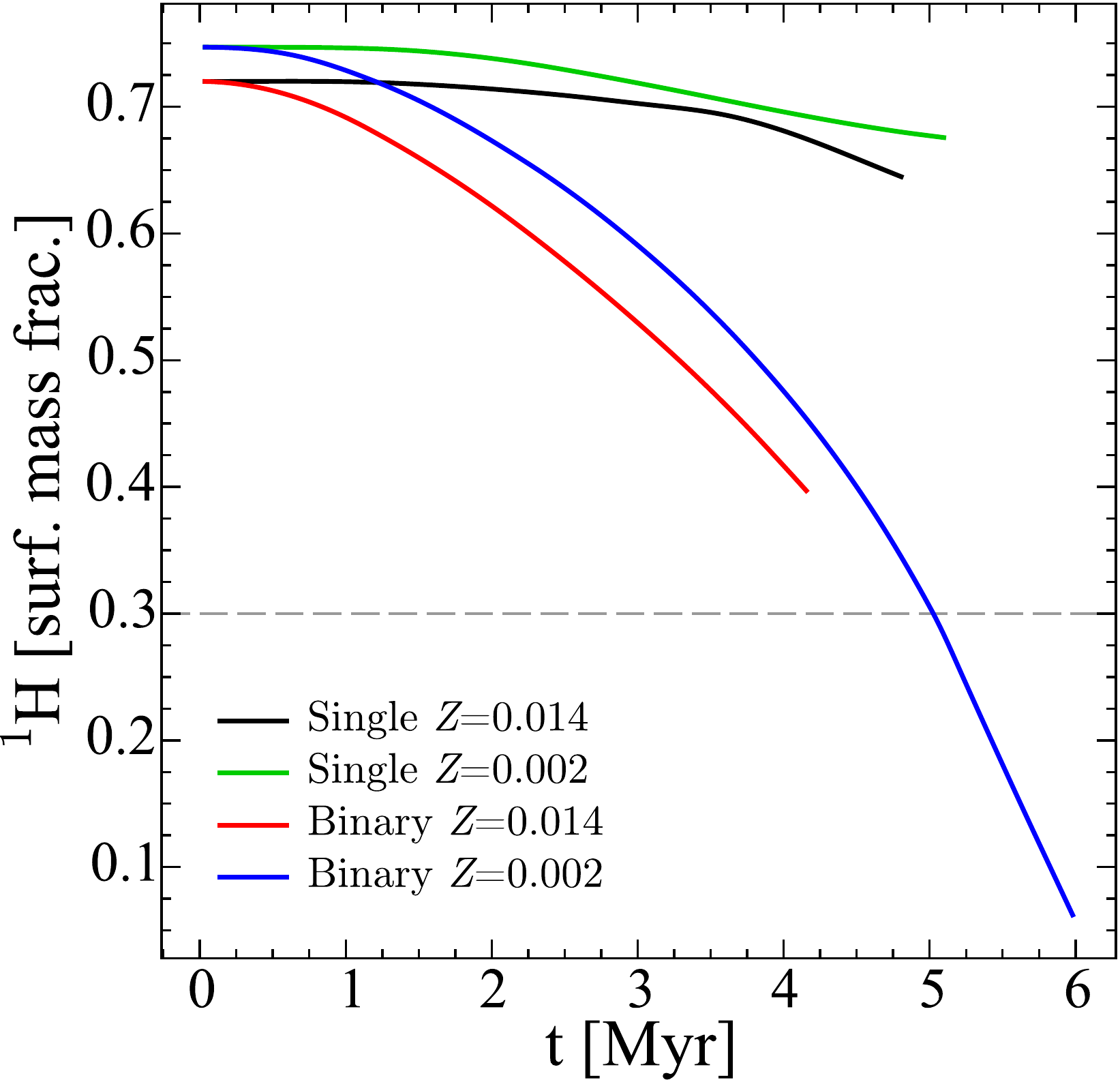}   
      \caption{Evolution of the surface hydrogen abundance (in mass fraction) for 40 M$_\odot$ models.
      The limit for becoming a WR star ($\mathrm{X_s}<0.3$) is indicated with a grey dotted line. The models plotted are the same
      as those presented in Figs.~\ref{m40sd} and \ref{m40su}.
       {\it Left panel:} Cases of spin-down. The $Z$=0.014 and $Z$=0.002 models have initial rotation on the ZAMS equal to respectively 365 km s$^{-1}$ and 391-394 km s$^{-1}$ (see Table.~1)).  {\it Right panel:} Cases of spin-up. The $Z$=0.014 and $Z$=0.002 models have initial rotation on the ZAMS equal to respectively 151 km s$^{-1}$ and 163-167 km s$^{-1}$.}
         \label{M40WR}
   \end{figure*}

In Fig.~\ref{M40WR}, the evolution of the mass fraction of hydrogen at the surface of the 40 M$_\odot$ stellar models are presented for the metallicities $Z$=0.014 and 0.002.
The single star models starting with a larger initial rotation rate (see black and green curves in the left panel) show stronger depletion of hydrogen at the surface than those starting
with a lower initial rotation (see black and green curves in the right panel). We see in particular that the single star model at Z=0.002 starting with an initial rotation of 391 km $s^{-1}$
enter into the regime when the hydrogen mass fraction at the surface become inferior to 0.3. The star becomes a WR$_{e}$-type star (see Sect.~2).

\begin{figure*}
   \centering
    \includegraphics[width=8.9cm]{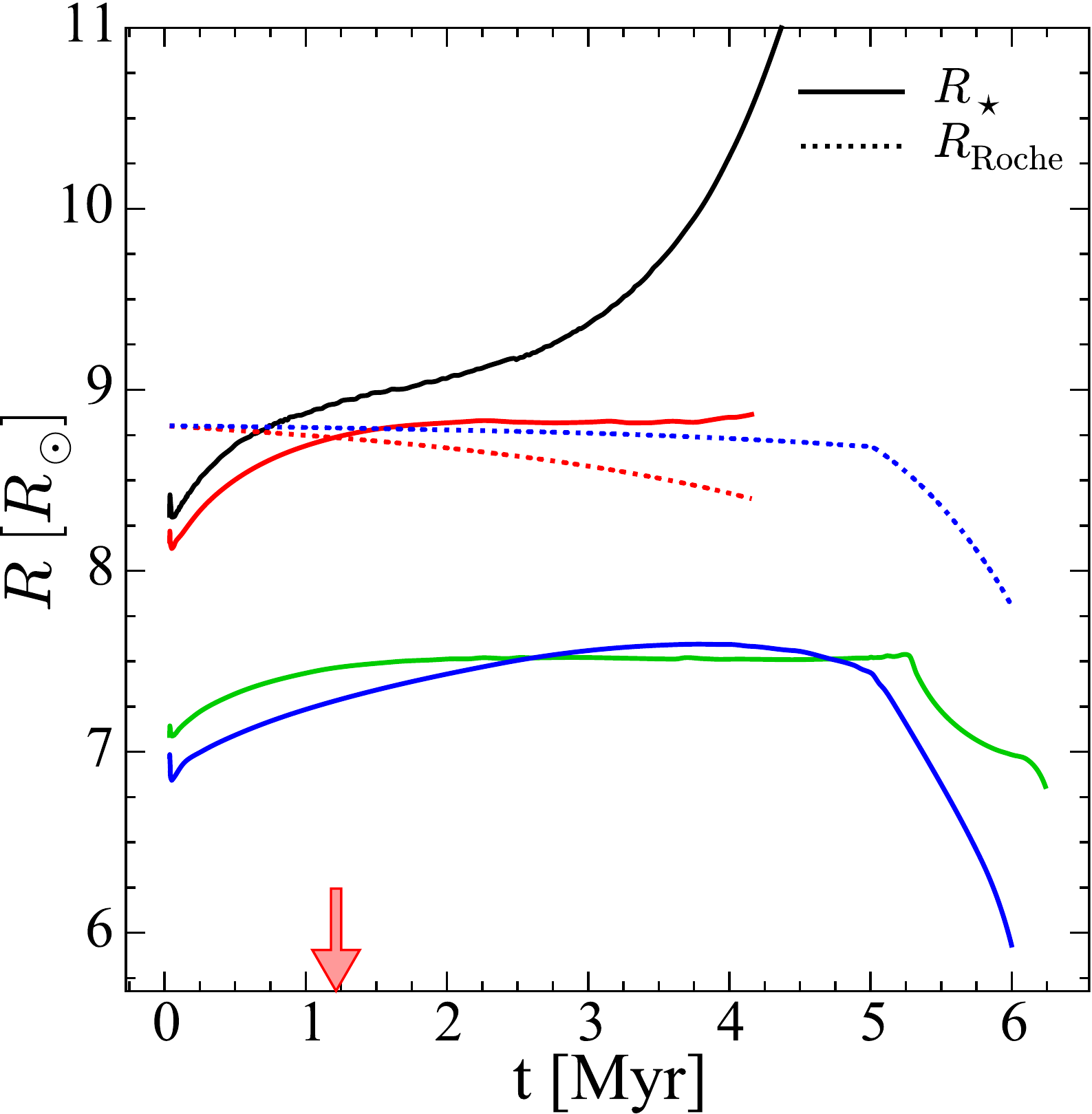}\hfill\hfill  \includegraphics[width=8.9cm]{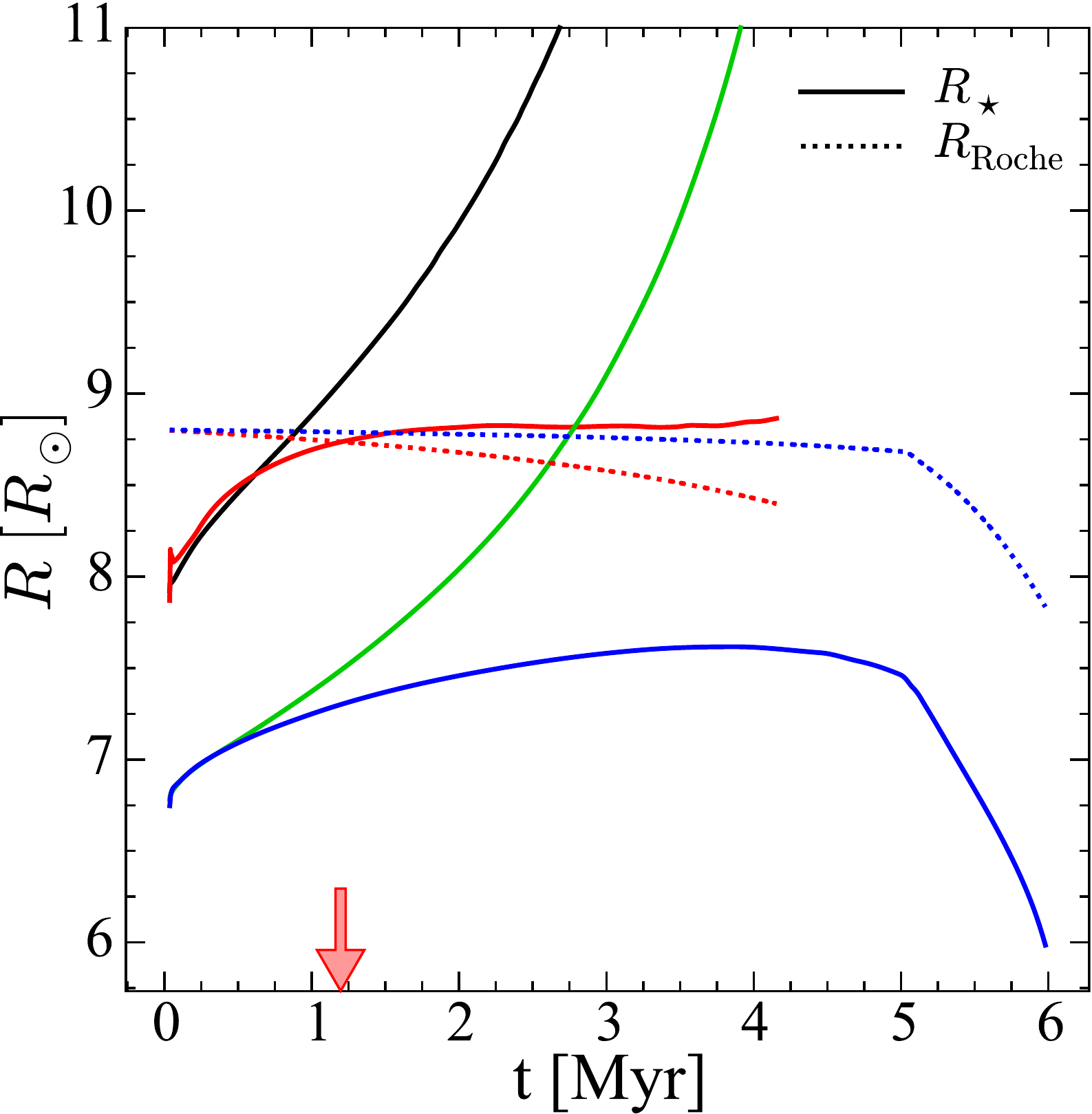}   
      \caption{Evolution of the stellar radius and the Roche radius for the same models as in Fig.~\ref{M40WR} (same color code). The time at which the Roche radius is reached by the Z=0.014 binary is marked with a red arrow.}
         \label{M40R}
   \end{figure*}

The tracks of the close binary models are quite similar for the case of spin-down and spin-up, which is expected since these two models reach very rapidly the same
velocities at the surface. We just note that the $Z$=0.014 model stops at an earlier phase
than the model at $Z$=0.002 since it encounters the Roche limit while the lower metallicity models avoids it. This can be seen in Fig.~\ref{M40R}, where the evolution as a function of time of the radii
of the 40 M$_\odot$ models are shown. The $Z$=0.014 close binary model  crosses the Roche limit at an age equal to 1.2 Myr (both in case of spin-down and spin-up), while due to strong mixing, the
$Z$=0.002 models never crosses it during the whole MS phase.

\section{The whole grid of stellar models}



\begin{figure*}
   \centering
    \includegraphics[width=18cm]{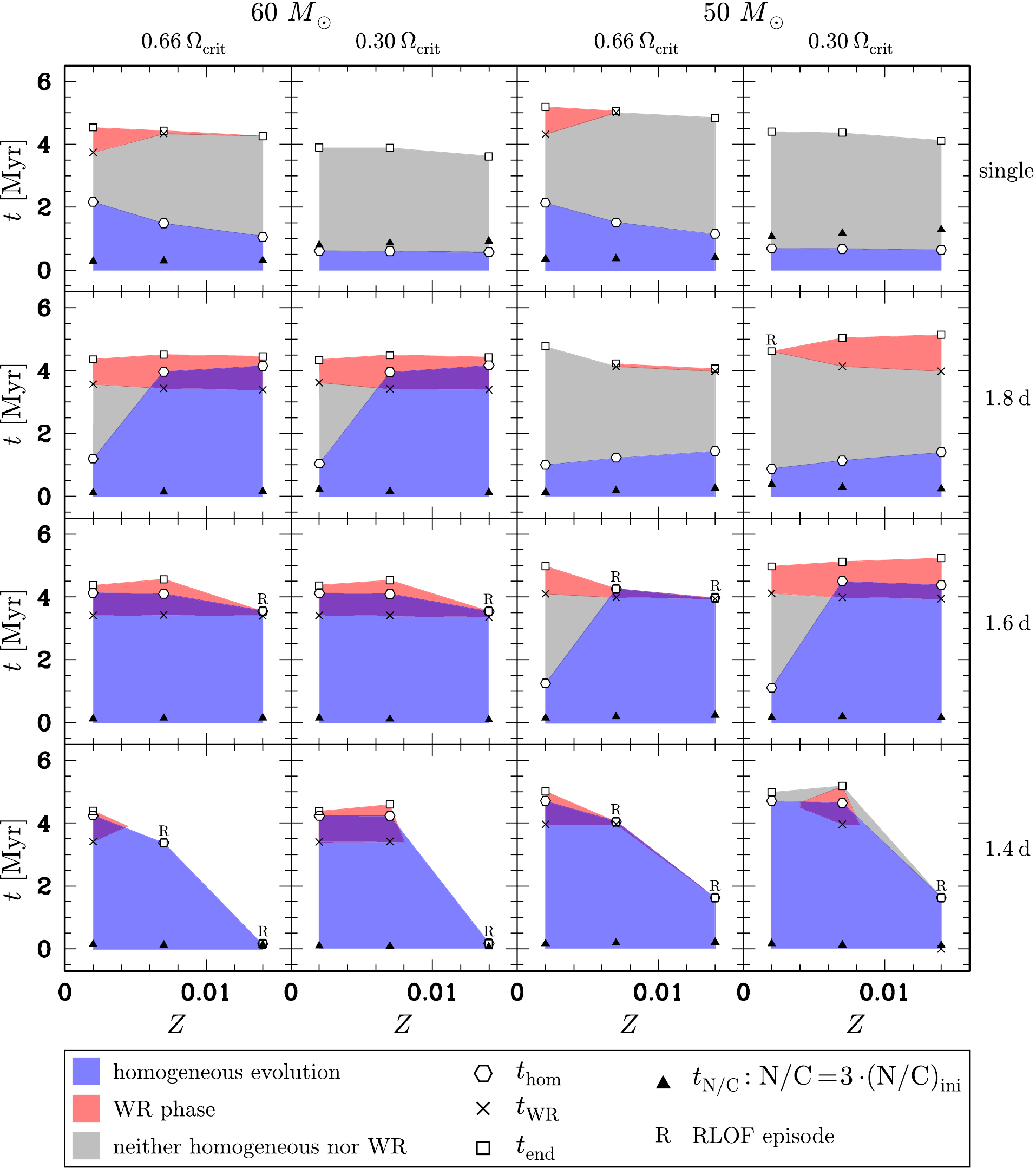}     
      \caption{For each stellar model, with initial masses equal to 60 and 50 M$_\odot$,  
      $t_{N/C}$ (black triangles), $t_{\rm hom}$ (empty hexagons),  $t_{\rm WR}$ (crosses) and $t({\rm end})$,  the age of the last computed model (empty squares), are indicated as
      a function of metallicity. These times are defined in the text in Sect.~3. The blue area corresponds to the part of the evolution which is homogeneous, the red region
      to the WR phase. When the stellar models are neither homogeneous, nor a WR star, the region is in grey. When a R is indicated, it means that the model reaches a stage where
      Roche lobe overflow occurs.}
         \label{time6050}
   \end{figure*}

\begin{figure*}
   \centering
    \includegraphics[width=18cm]{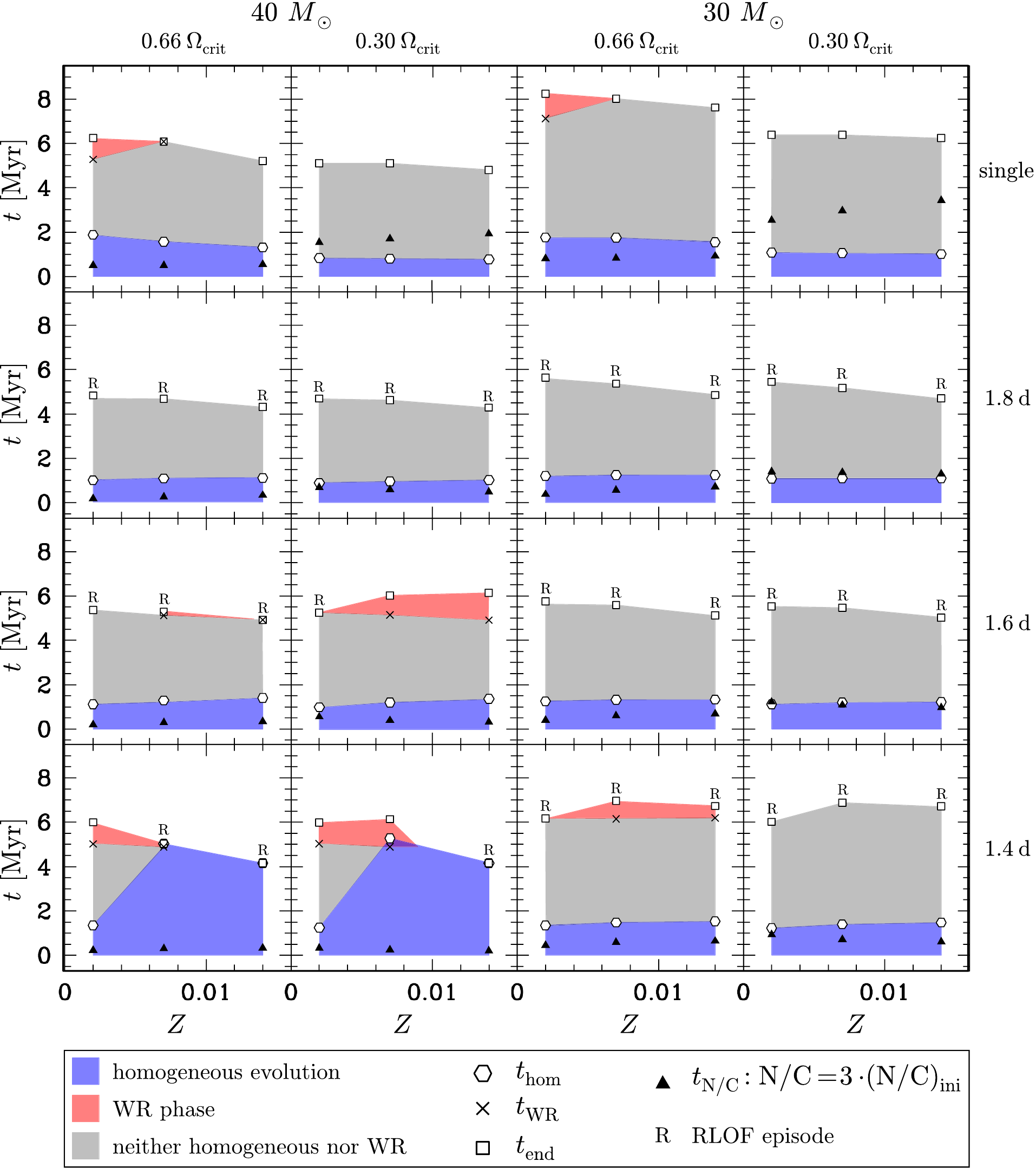}     
      \caption{Same as Fig.~\ref{time6050} for initial masses equal to 40 and 30 M$_\odot$.}
         \label{time4030}
   \end{figure*}

To discuss the impact of the synchronization process on our stellar models, we introduce below a few specific times:
\begin{itemize}
\item $t_{\rm sync}$, the synchronization timescale,  defined here as the time for the surface angular velocity to approach the orbital angular velocity to less than 10\%;
\item $t_{\rm N/C}$, the age of the star when, for the first time the ratio of nitrogen to carbon at the surface becomes larger than three times the initial ratio (the initial ratio is equal to 0.29 in mass fraction);
\item $t_{\rm hom}$, the age of the star when the difference between the surface  and central hydrogen mass fraction becomes larger than 10\% the central hydrogen mass fraction. Even in classical (non-rotating) models,
with no extra mixing, there is always a small period during which the mass faction of hydrogen at the surface is equal or near to the mass fraction at the center since stars on the ZAMS are homogeneous and it takes
some time for the central hydrogen to be depleted by more than 10\%. Of course when mixing is active, this time is increased.  {\it To some extent} this time indicates the duration of the homogeneously evolving
part of the stellar lifetime (indicated by the blue shaded regions in Figs~\ref{time6050} and \ref{time4030}). 
We say {\it to some extent}, because the evolutionary track in the HR diagram can present the characteristics of a homogeneous evolution even when the difference between the surface and central hydrogen mass fraction is larger than 10\%, since
homogeneous evolution depends
also on the way hydrogen is distributed {\it inside} the model and not only on the difference between the mass fraction of hydrogen
at the surface and at the center. 
\item $t_{\rm WR}$ is the age of the star when it enters the Wolf-Rayet (WR) phase (see Section~2).
\item  $t({\rm end})$ is the age of the last computed model. This age corresponds approximately to the evolutionary stage when the primary fills its Roche Volume (when this occurs) or to the end of the Main-Sequence phase (when the filling of the Roche volume  does not occur).
\end{itemize}

Figures~\ref{time6050} and  \ref{time4030} show how the various times  change as a function of the initial metallicity for different initial masses, rotations and orbital periods (for binary systems)\footnote{The reader will find in the Appendix A of this paper,  tables with various properties of all the models computed here, among them the precise values of the times defined just above.}.
The synchronization timescale is not shown since it is too short (shorter than $t_{N/C}$).
These figures allow to see the impact of the tidal forces
for various initial masses, rotations and metallicities,
by comparing the upper panels corresponding to single star evolution and the lower panels showing the situation when a companion is present.
For each initial mass, the figures on the left column show the cases of spin-down, while the figures on the right column show the cases of spin-up.


\subsection{Models of single stars}

Before to discuss the effects of the tidal interactions, let us discuss the results obtained for single star models or for stars in wide binary systems.
In our previous works on rotating models, we showed that
rotating models computed with no magnetic field present the following behavior for what concerns the efficiency of the rotational mixing:
this efficiency increases with the initial mass, the initial rotation and decreases with the initial metallicity  \citep[see e.g. the discussions in][]{MM01,Ekstrom12, Georgy13a}.
From previous work, we know that  models with an internal magnetic field are more mixed
\citep{MM05}, but we have never investigated how mixing triggered by magnetic instabilities behaves when the initial mass, rotation
or metallicity vary. {\it Actually, it happens that these dependencies are exactly  the same (qualitatively) than those that were deduced from rotating non-magnetic models}. Let us see why in the following.

Mixing in magnetic models is more efficient in higher mass stars. 
Comparing for instance the positions of the triangles in the upper panels of Fig.~\ref{time4030} corresponding to an initial rotation equal to about 30\% the critical angular velocity for a 40 and  30 M$_\odot$ model (the second and the fourth panel starting from left), we see that the stage at which the N/C ratio at the surface becomes superior to three times
the initial one corresponds to about half the main sequence lifetime in the case of the 30 M$_\odot$ at Z=0.014, while it corresponds to about one third of the MS lifetime for the corresponding model with an initial mass equal to 40 M$_\odot$. This dependence on the initial mass is such because
$U$, the amplitude of the radial component of the meridional velocity, which is the main driving parameter 
for the chemical mixing in magnetic models, scales with the luminosity to mass ratio \citep[see Eq. 11.75 in][]{Maeder09}, a ratio which increases
when the initial mass increases.

As expected, mixing is more efficient in faster rotating models.
Typically compare the positions of the triangles in Fig.~\ref{time4030} in the upper two right panels corresponding to 
single 30 M$_\odot$ models. We see that when the initial rotation increases, triangles are shifted to smaller times 
indicating that
in faster rotating models, mixing is more efficient.
This comes from the fact that $U$
scales with $\Omega$.

At higher metallicities, the stellar models are less efficiently mixed. Again this can be observed by looking at the triangle positions for instance in the upper right panel
of Fig.~\ref{time4030}. We see that for the spin-up 30 M$_\odot$ stellar model, the time at which N/C at the surface becomes larger than three times  the initial ratio is shifted towards larger ages when higher metallicities are considered.
This comes mainly from the fact that,
at high metallicities, stars are efficiently slowed down by the loss of angular momentum by stellar winds and thus are less mixed.
In models with no strong stellar winds, two counteracting effects intervene that may, depending on the circumstances,
trigger more or less mixing at high than at low metallicity.
A rough estimate of the timescale for the mixing in the radiative zone can be obtained by the ratio 
of $U$ to the radius of the star  $R$
\citep[see Eq. 11.63 in][]{Maeder09}. 
On one hand, at higher metallicities, $U$
is larger in the outer layers (since $U$ scales as the inverse of the density in this region and stars, in metal-rich regions, are
larger and hence less dense in their outer layers than stars in metal-poor regions). On the other hand,  $R$ is larger at high metallicity. So the ratio $U/R$ may increase/decrease
with the metallicity.
In the present paper, for the mass range considered, the effect of the mass loss dominates making the mixing
more efficient in low metallicity stars.

\begin{figure*}
   \centering
    \includegraphics[width=8.5cm, angle=0]{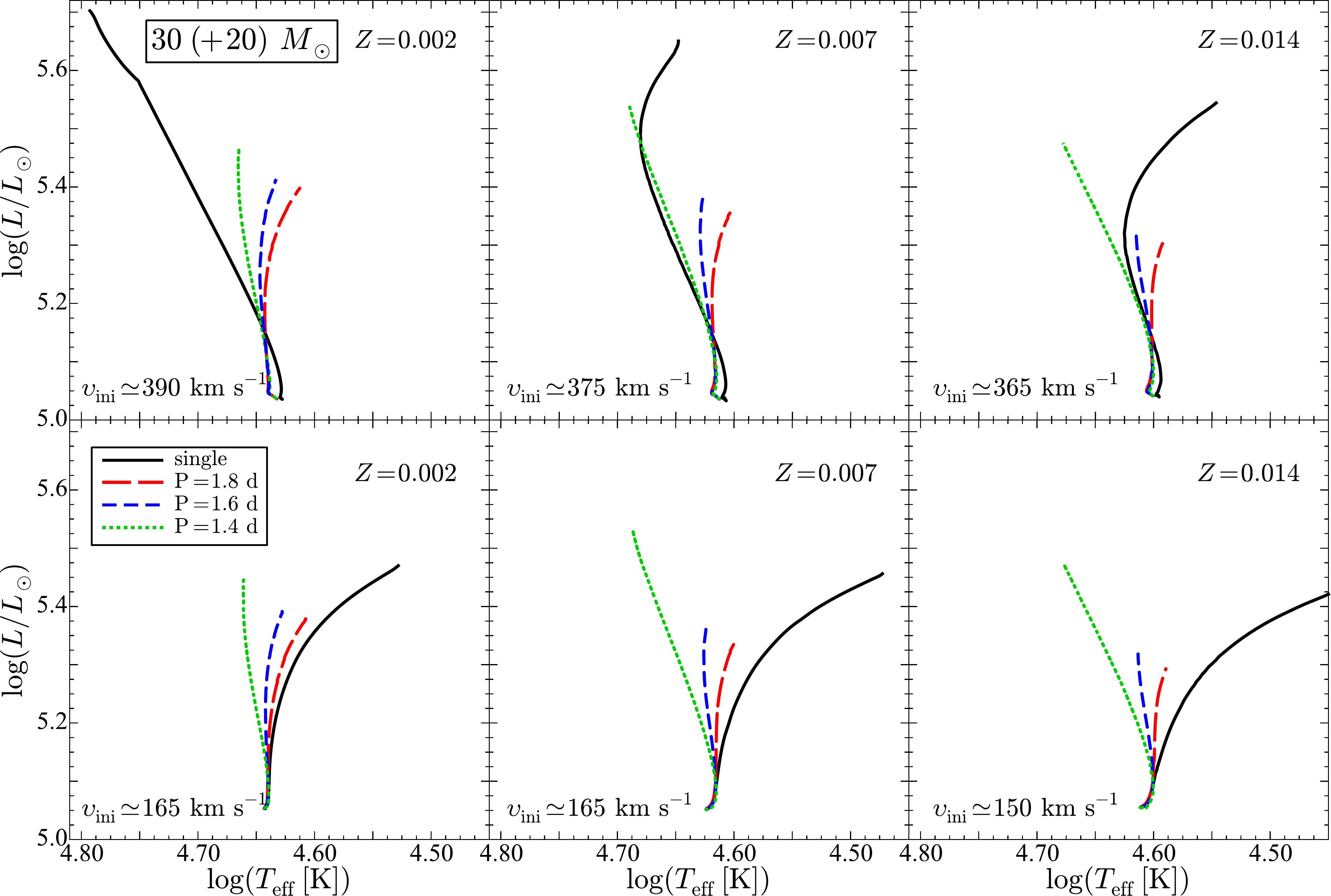}  \includegraphics[width=8.5cm, angle=0]{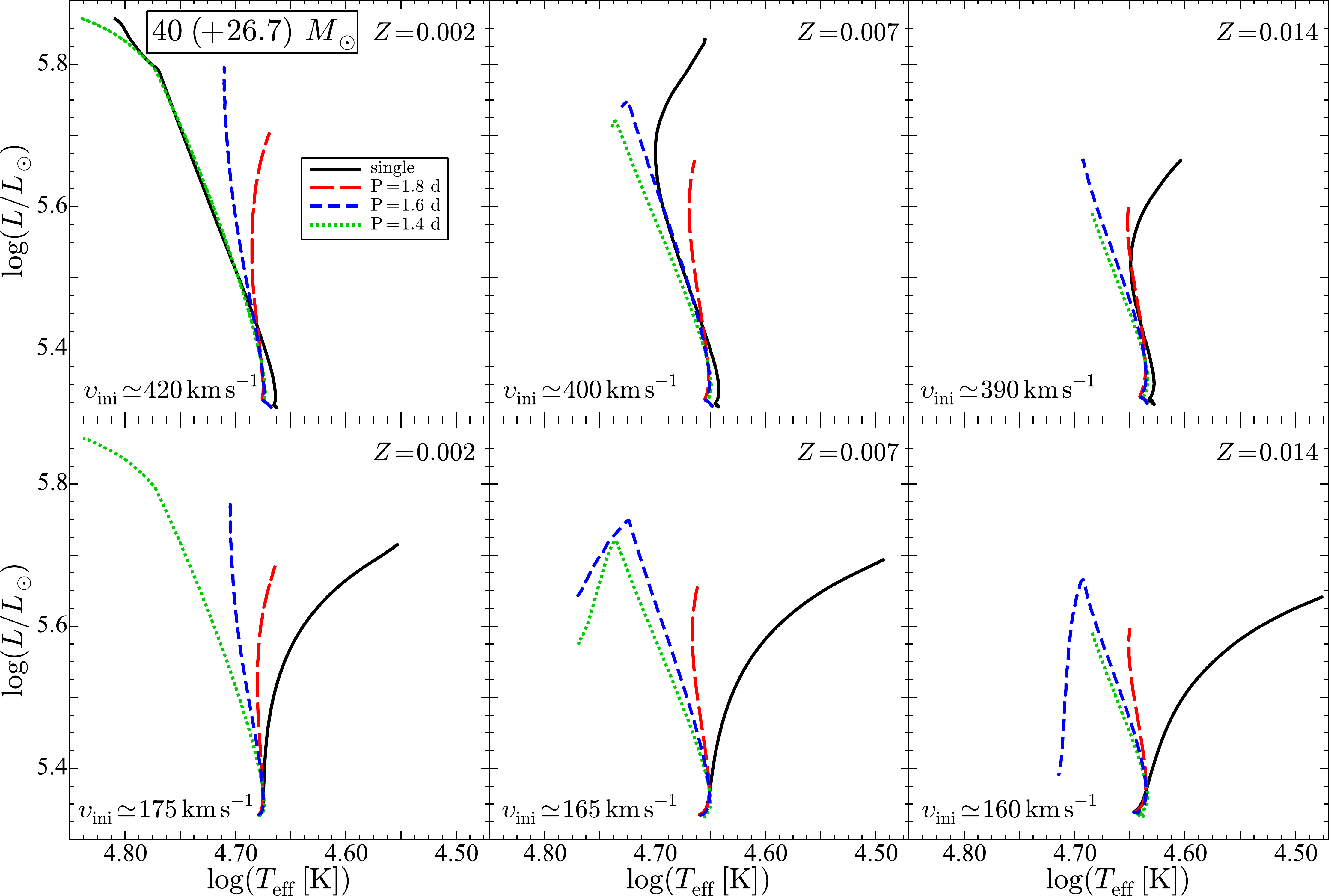}   
    \includegraphics[width=8.5cm, angle=0]{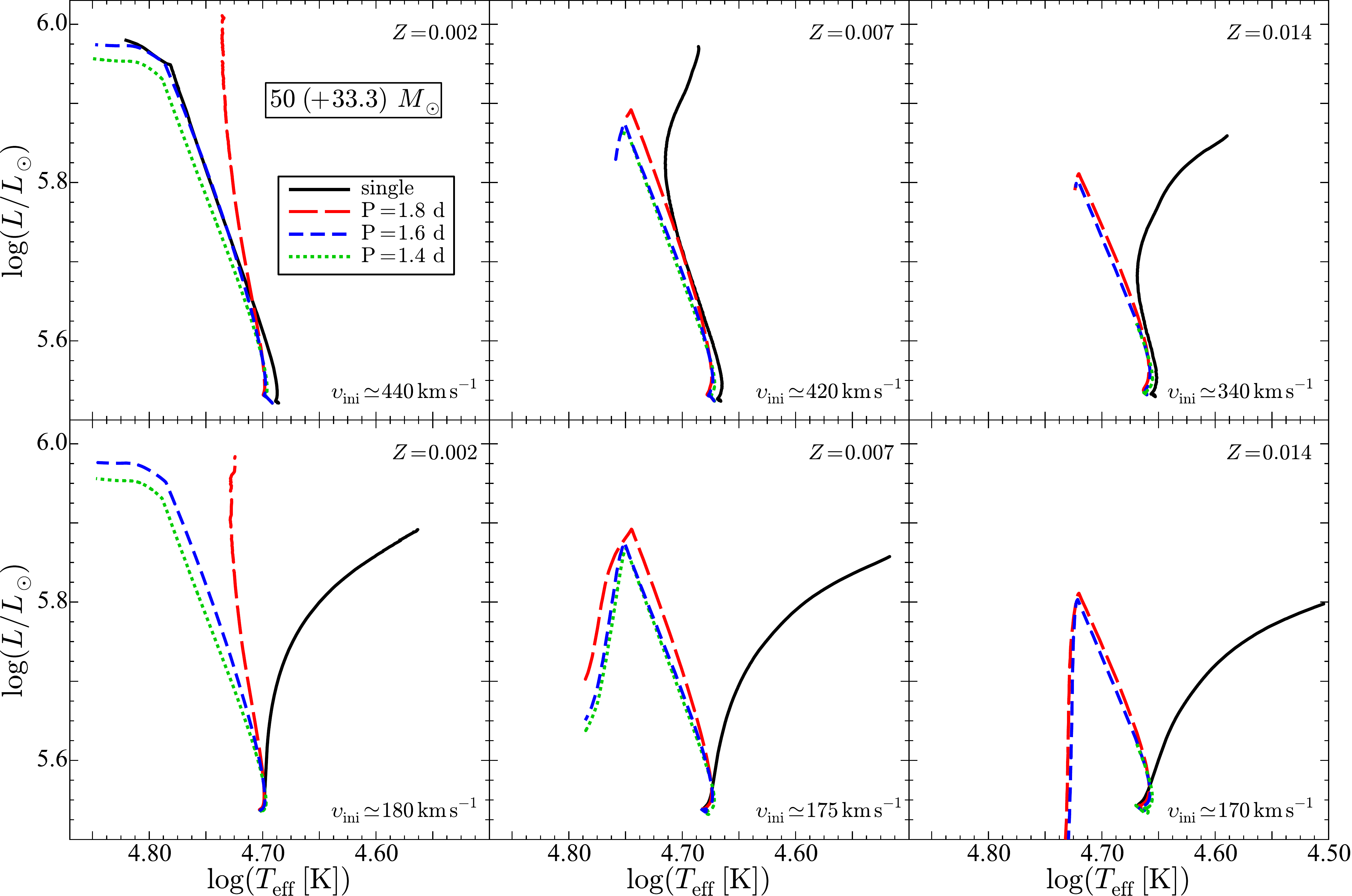}  \includegraphics[width=8.5cm, angle=0]{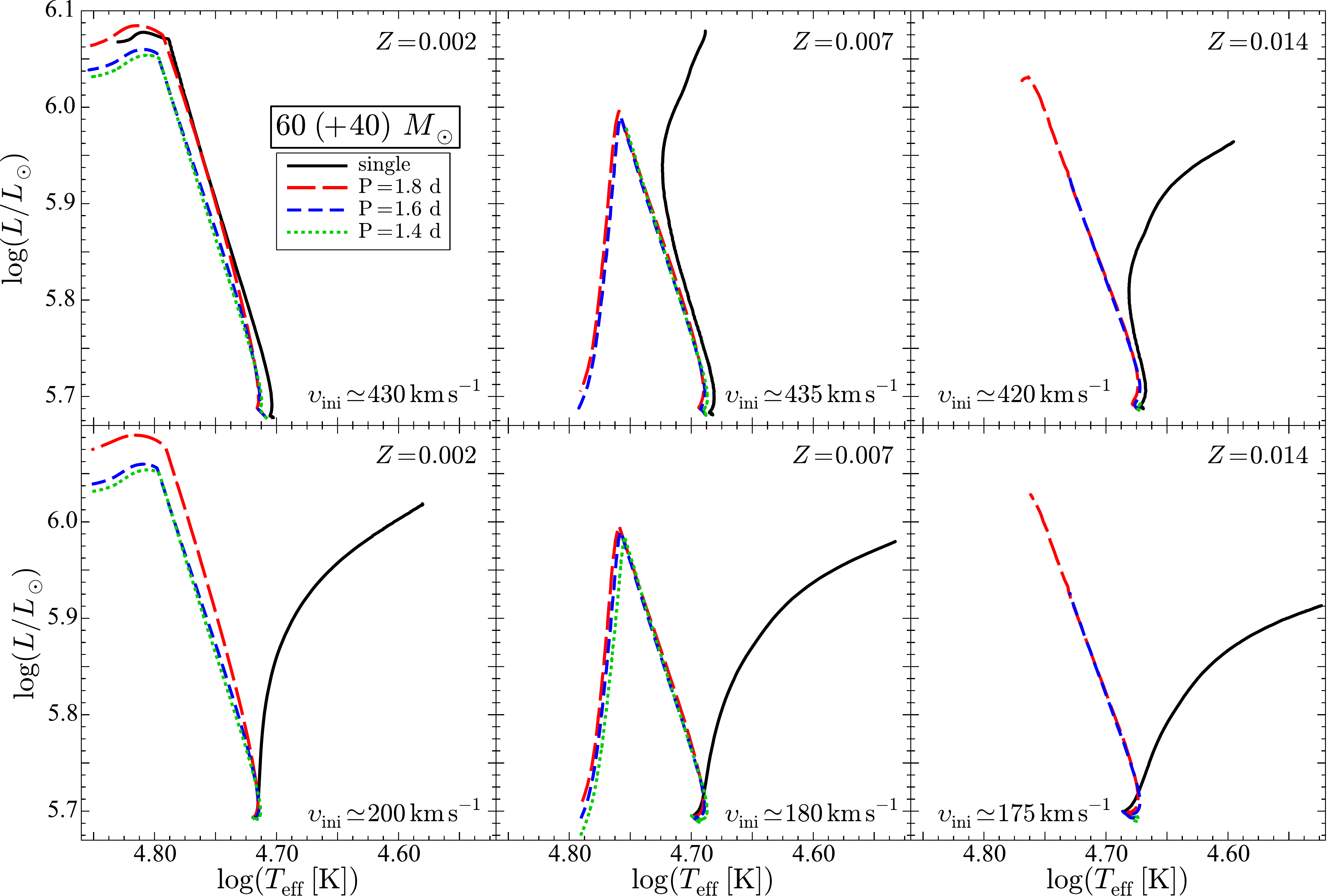}       
      \caption{Evolutionary tracks in the HR diagram for single star models and close binaries.
      The four panels correspond to different initial masses for the single star or for the primary in the close binary (in that case, the secondary has a mass equal to 2/3 the primary mass). The masses, the initial metallicities (Z) and
      the approximate initial surface velocities on the ZAMS are indicated. For the precise initial surface rotation of each model look at the values in Tables A.1 and A.2. For a given initial mass, the three upper panels correspond
      to spin-down cases, while the three lower panels correspond to spin-up cases.
    }
         \label{alldhr}
   \end{figure*}

All the single star evolutionary tracks are shown in Fig.~\ref{alldhr} (look at the black lines). For all the initial masses considered, the faster rotating tracks
at Z=0.002 have a homogeneous evolution. When the metallicity increases, the fast rotating tracks (look at the upper panels for each mass) begin to bend redwards at still earlier stages.
The single star tracks computed for a moderate initial rotation (lower panels for each mass) show all a classical redwards evolution during the MS phase. When the metallicity increases,
the tracks are redder and less luminous.

\subsection{Models of close binary stars.}

We can note three types of main differences between the models with and without tidal interactions: changes in the surface
velocities and compositions, changes in the evolutionary tracks and possible occurrence of Roche lobe overflow event in close binary systems. 


\subsubsection{Synchronization times}

How does the synchronization time varies with the initial mass, metallicity, rotation and orbital period?
For the models shown in Tables A.1 and A.2, $t_{\rm sync}$ is always inferior to 0.8 Myr.
In general it is much shorter than this upper value and corresponds to a few percents of the MS-lifetimes.  

It increases with the difference between the initial $\Omega$ and the orbital period: see for instance the 30 M$_\odot$ at Z=0.014 starting with an initial equatorial rotation
velocity of 365 km s$^{-1}$, the synchronization time is 0.18 Myr when $\Omega/\omega_{\rm orb}$=1.4, and equal to 0.43 Myr when
$\Omega/\omega_{\rm orb}$=1.8 ($\Omega$ is the stellar surface angular velocity and $\omega_{\rm orb}$ is the orbital angular velocity).  

The dependency on the initial metallicity does appear quite weak. For instance
the model B1.6 for the 30 M$_\odot$ model that has an initial value of $\Omega/\omega_{\rm orb}$=1.773 at Z=0.007, has a value of $t_{\rm sync}$=0.36 Myr. The model B1.4 for the 30 M$_\odot$ model that has an initial value of $\Omega/\omega_{\rm orb}$=1.772 at Z=0.002, has a value of $t_{\rm sync}$=0.34 Myr, so not significantly different. 

The synchronization timescale decreases when larger initial masses are considered: compare for instance $t_{\rm sync}$ for the 30 M$_\odot$
at $Z=0.007$ with $\Omega/\omega_{\rm orb}$=1.551, that is equal to 0.23 Myr, with the corresponding $t_{\rm sync}$ value for the
60 M$_\odot$ at Z=0.007 with $\Omega/\omega_{\rm orb}$=1.557, that is equal to 0.08. Note that the MS lifetime for the 60  M$_\odot$
is about 55\% the MS lifetime of the 30 M$_\odot$, while $t_{\rm sync}$ for the 60 M$_\odot$ is about one third the value of $t_{\rm sync}$ for the
30 M$_\odot$, indicating that $t_{\rm sync}$ normalized to the MS-lifetime decreases when higher initial mass stars are considered.

\subsubsection{Surface and internal rotation rates}

\begin{figure*}
   \centering
    \includegraphics[width=18cm]{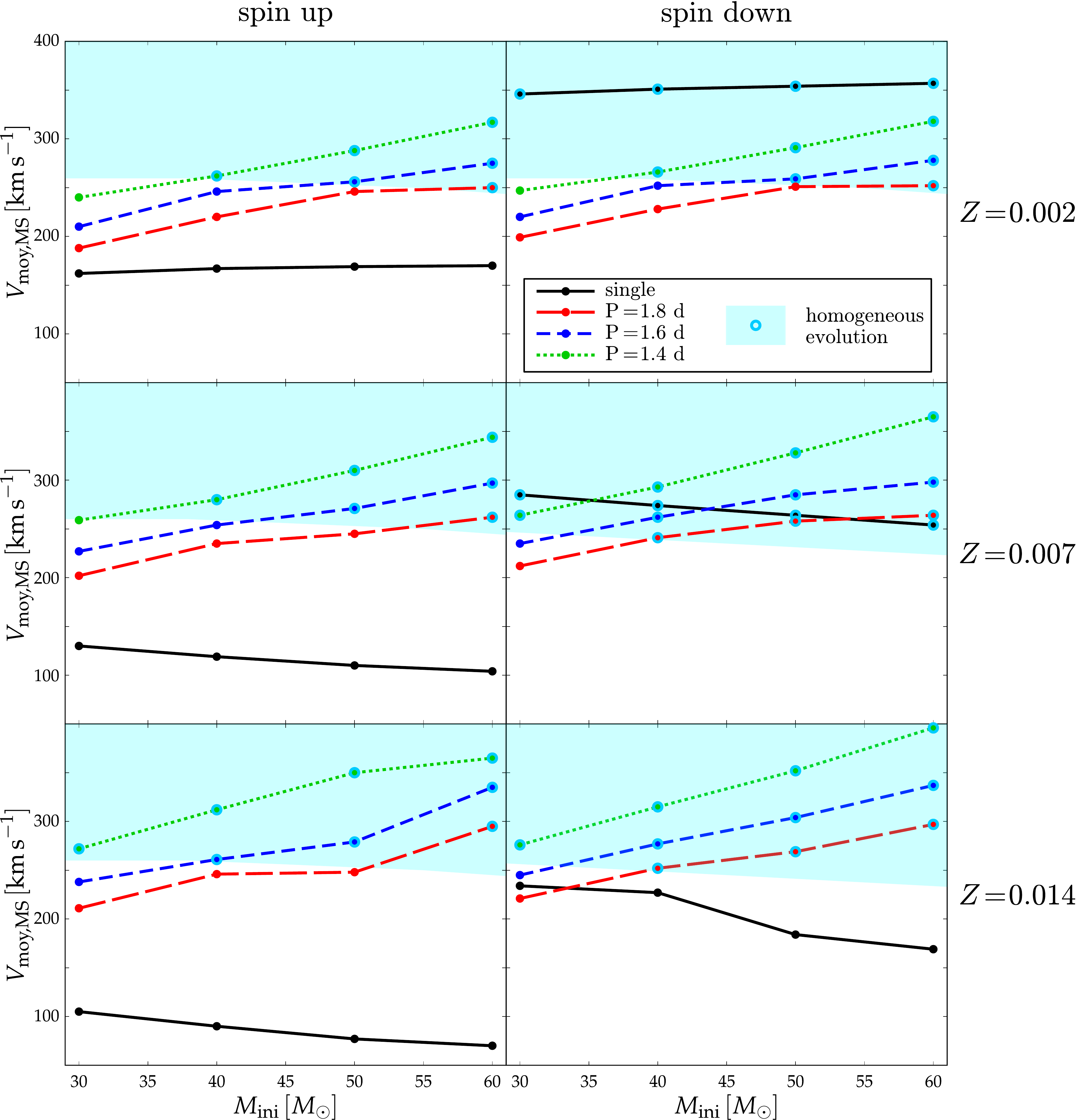}     
      \caption{Time-averaged surface equatorial velocity of single and binary stars as a function of the initial mass for various metallicities. The (black) continuous line connects the single star models. The (colored) dashed lines connect the binary stellar models.
      The initial period in days is indicated. The results corresponds to a companion having 2/3 the mass of the primary. The (blue) big dots show the models that follow a homogeneous evolution. The zone where homogeneous models lay in these diagram is
      hatched in blue. The panels on the left column correspond to spin-down cases and the panels on the right column to spin-up cases.}
         \label{veqsdsu}
   \end{figure*}


Time-averaged surface velocities for all the models as a function of the initial mass, metallicity and orbital period are shown in Fig.~\ref{veqsdsu}. The curves labeled by an S show the cases of single star models or models of stars belonging to wide binary systems.
The curves labeled with a value of an orbital period correspond to close binary models.
The most remarkable points to be noticed are the followings:
\begin{itemize}
\item For single stars, we clearly see the impact of stronger mass losses for higher metallicities and higher stellar masses. This makes
the time-averaged surface velocities to decrease when the metallicity increases and when the initial mass increases.
\item In case of close binary stellar models which are spin-down by tidal interactions (see the left panels), we see that at low metallicity,
the spin-down is well effective. All the surface velocities in the close binary systems are smaller than the surface velocities of the
single star models. It is quite different at higher metallicities. For instance at $Z=0.014$, the surface velocities in the close binary systems are larger than the surface velocities of the
single star models. This might appear strange, since at the beginning we have a situation of spin down. What does happen here is that
the spin down does occurs only up to the point when synchronization is achieved. Once synchronization  is reached, the transfer of angular momentum
from the orbit to the star compensates for the losses of angular momentum by the star due to stellar winds. Thus we have actually from that point
an acceleration process which sets in.
\item  We see that the time-averaged velocities in the close binary systems are slightly larger at the higher than at lower metallicities. This comes from the fact that
stars at high metallicities are larger than at low metallicities, implying that for a given surface angular velocity, imposed by synchronization, the surface linear
equatorial velocities are larger at high than at low metallicity\footnote{the angular surface velocity is determined by synchronization and thus depends only on the orbital period, but not 
on the initial metallicity.}.
\item Globally the cases of spin-up and spin-down are very similar. The largest differences occur for the 60 M$_\odot$ at $Z=0.014$ for a period equal to 1.4 days, the time-averaged
surface velocity is equal to 400 km s$^{-1}$ in the case of spin-down and to 370 km s$^{-1}$ in case of spin-up. This comes from the fact that the synchronization is obtained more rapidly
in the spin-down case than in the spin-up case because the initial rotation, in the spin down case is much nearer from the synchronized one than in the case of spin-down (see Table A.2).
 \end{itemize}

All the models with tidal interactions show a very low degree of contrast between the central and the surface angular velocity.


\subsubsection{Evolutionary tracks}


All the tracks for the primaries in various close binary systems are shown in Fig.~\ref{alldhr}. For all the initial masses, the close binary tracks
will be comprised between the following two extreme cases: the single fastest rotating track at Z=0.002 and the single slowest rotating track at Z=0.014.
We see also that the close binary tracks show great similarities for the spin-down (upper panels for each mass) and the spin-up (lower panels for each mass),
which is indeed expected (this is particularly visible in the case of the 30 M$_\odot$).

In the left panel of Fig.~\ref{synt}, we indicate the models that follow a homogeneous evolution. We used a very strict and conservative criteria for considering that a star follows a homogeneous evolution, namely, only those
models evolving blueward along a straight line (until they enter into a WR phase) are considered to evolve homogeneously.
There is no difference between the spin-down and spin-up cases.
For each metallicity and initial mass, there exists an upper value for the orbital period below which stars evolve homogeneously.
These upper limits are indicated in Table~\ref{tabH}.
We see that the conditions for having a homogeneous evolution in close binary systems are more restricted
at low than at high metallicities (the upper limits for the periods are smaller at Z=0.002 than at Z=0.014).
We explain the reason for this in the next subsection.

All computations have been performed for a single value of the mass ratio between the secondary  and the primary equal to 2/3. One can wonder
how things change when this ratio varies. For a given mass of the primary and a given initial orbital period, a higher mass ratio implies stronger tides and thus a more rapid synchronization time. Note that
keeping the period constant, assuming a more massive secondary star implies that the distance between the two components is larger, thus, a priori, it is
not obvious that tides are stronger. This can however be seen using the third Kepler's law and equation (6) in paper I for the synchronization time.
Thus a higher mass ratio favors homogeneous evolution through the decrease of the synchronization time\footnote{Note that the angular velocity of the stars at synchronization is
fixed by the orbital period and thus, provided this quantity is kept fixed, the mass ratio has no impact on the post-synchronization angular velocity.}.

\begin{table}
\caption{Maximum initial periods, $P_{\rm max}$, in days for having a homogeneous evolution for different initial
mass stars (in solar masses) and metallicities and for a mass ratio of 3/2. When  $P_{\rm max}$ is larger than the largest period, respectively, smaller than the smallest period explored in this work,
the symbols $>$ and $<$ are used.}
 \label{tabH} 
\begin{tabular}{cccc}
Mass & Z=0.002 & Z=0.007 & Z=0.014 \\
\hline\hline
& & &  \\
60            &  $>$ 1.8    &     $>$ 1.8    &   $>$ 1.8   \\
50            &   1.7        &    $>$ 1.8     &  $>$ 1.8  \\
40            &   1.5         &  1.7         &   1.7   \\
30            &   $<$1.4      & 1.5          &   1.5    \\
& & &  \\
\hline\hline
\end{tabular}
\end{table}


\subsubsection{Mixing and Roche lobe overflow during the MS phase}

\begin{figure*}
   \centering
    \includegraphics[width=8.8cm, angle=0]{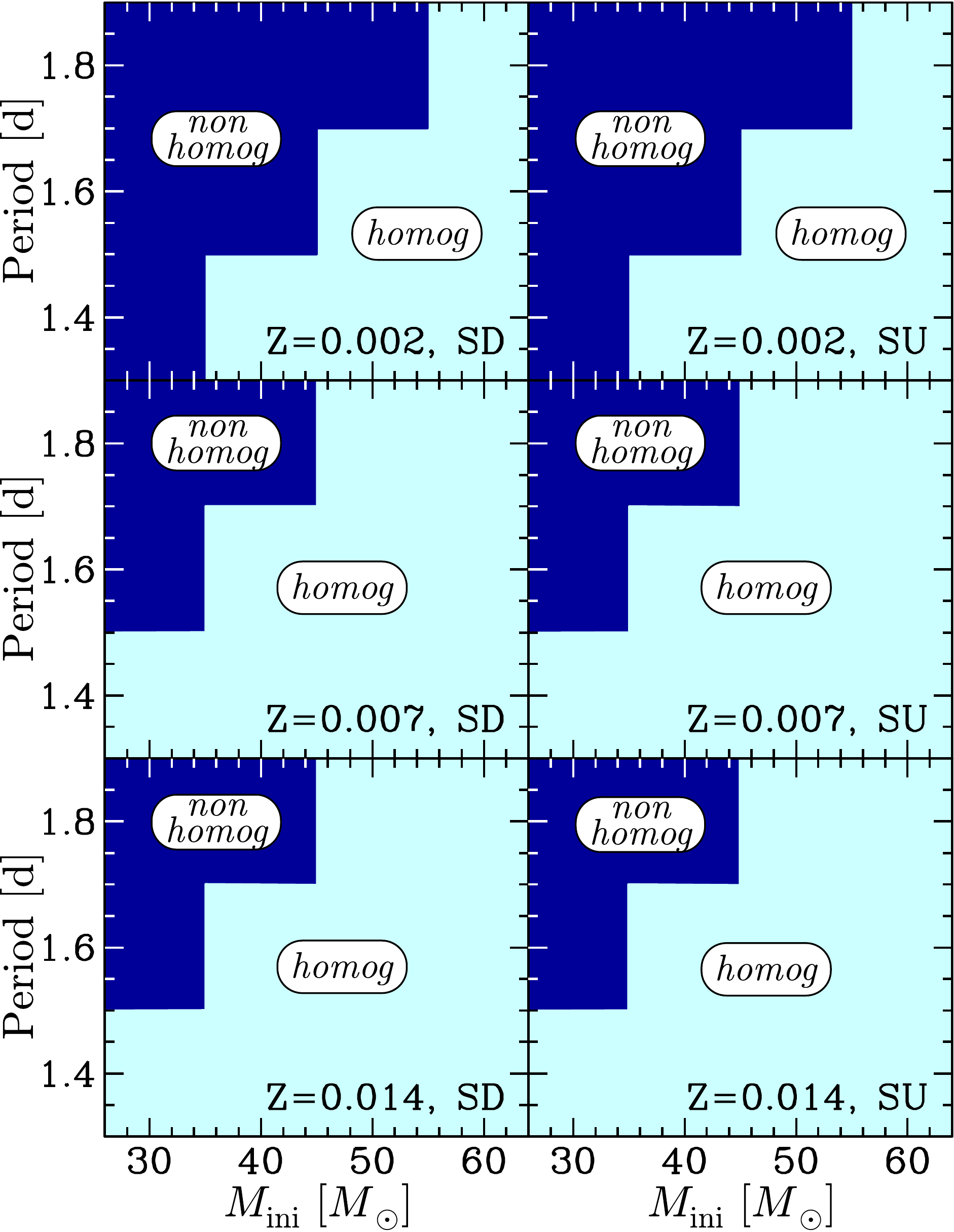}  \includegraphics[width=8.8cm, angle=0]{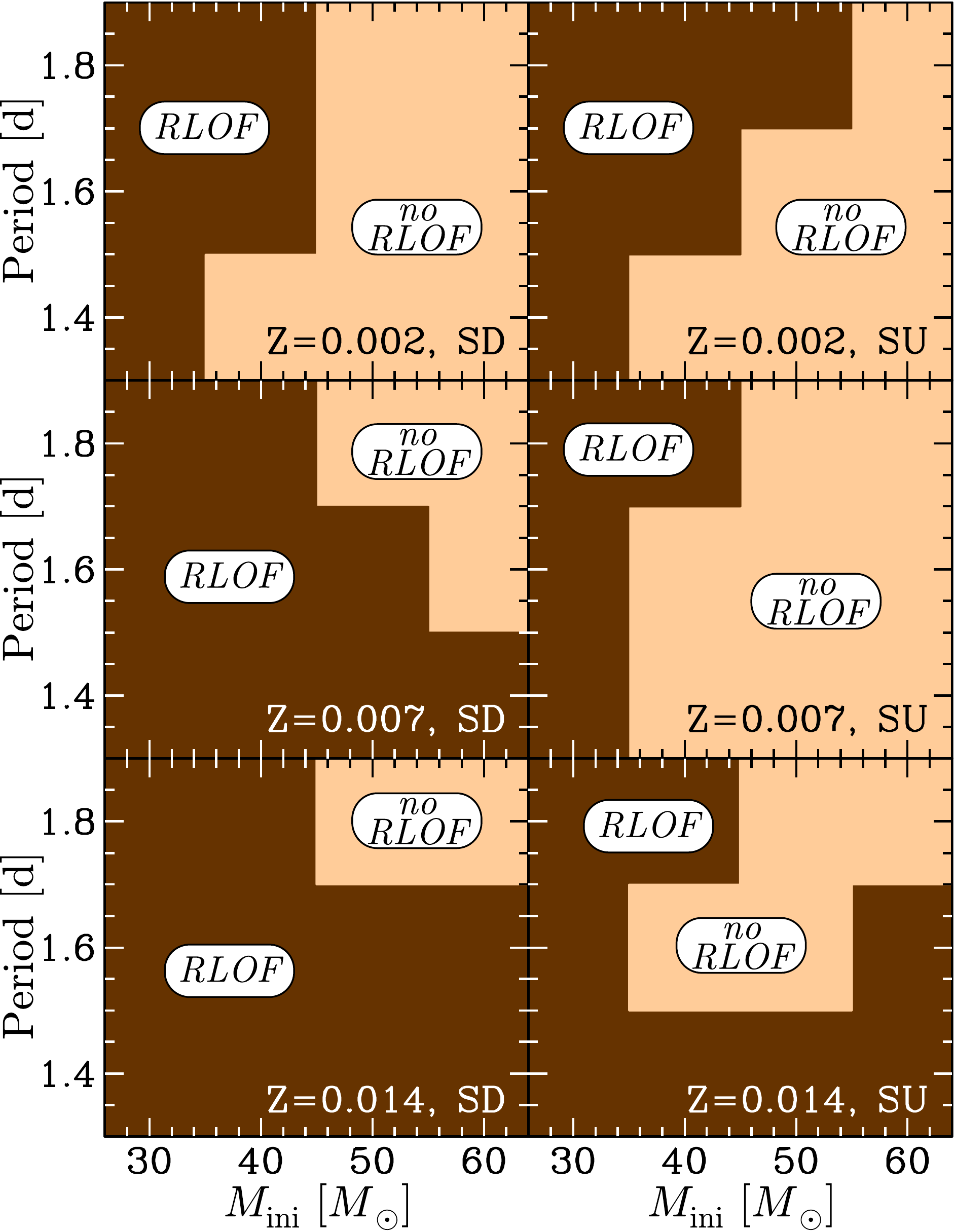}   
      \caption{
The vertical axis is the initial orbital period in days, the horizontal axis is the initial mass of the primary in a close binary system.
The secondary has a mass equal to 2/3 the mass of the primary. The different panels correspond to different initial metallicities $Z$
and to different initial rotation of the primary. The letters SD are for Spin-Down cases. The letters SU are for Spin-Up cases.
{\it Left panel:}  A homogeneous evolution  during the Main-Sequence phase occurs where {\it homog} is indicated (light blue areas).
Non-homogeneous evolution  during the Main-Sequence phase occurs where {\it non homog} is indicated (dark blue areas).
{\it Right panel:}  Roche lobe overflow during the Main-Sequence phase is avoided where {\it no RLOF} is indicated (light brown areas).
Roche lobe overflow during the Main-Sequence phase occurs where {\it RLOF} is indicated (dark brown areas).
    }
         \label{synt}
   \end{figure*}



Let us begin by discussing the cases of the 30 M$_\odot$ models in Fig.~\ref{time4030}.
Looking at the single models (see right upper panels of Fig.~\ref{time4030}), we recall the following main features: mixing is more important
when the initial rotation is larger and the metallicity smaller.
We see also that, at low metallicity, the fast rotating model enters the Wolf-Rayet phase at the end of the MS phase.

Now, when a companion is present with an orbital period comprised between 1.4 and 1.8 days, 
both in the spin-down and in the spin-up models, stars are more mixed 
(see for instance how the triangles are shifted to smaller values in binary models). We even see that the initially fast rotating model which is spin down (lower panel in the third column) enters the WR stage at the very end of the 
MS phase at Z=0.014, while the single corresponding model does not. As explained above this is due to the fact that, although this model is in a first phase spin-down, it will be actually spun up after synchronization in order to keep the surface spin angular velocity equal to the orbital angular velocity. This maintains a larger rotation than in the single star model which is continuously spun down by stellar winds. 
We see that for all these models, Roche lobe overflow
will occur before the end of the MS phase. 
Therefore for these models, we have never a situation where the Roche lobe overflow can be avoided during the MS phase.

Let us now discuss the case of the 60 M$_\odot$ models. The same qualitative features as for the 30 M$_\odot$ models can be seen for the plots representing the single star models. We just note that these more massive models are more mixed than the corresponding 30 M$_\odot$ models.
For binary models, we note the following striking features:

\begin{itemize}
\item As in the case of the 30 M$_\odot$ model, 
stars with tidal interactions are more mixed, whatever they belong to a system where a spin-down or a spin-up occurs.
\item One of the most striking point is the fact that mixing does appear to be more efficient in binary system at higher than at lower metallicity, in strong contrast with what happens for single stars (look for instance to the panels for the 60 M$_\odot$ models for an orbital period equal to 1.8 days, This is less visible for the shorter periods because the models meet the Roche limit). We explain this change of behavior in the following way: 
in the present models, the driving effects for chemical mixing are the meridional currents. It is well known that the meridional velocity in the outer
layers varies as the inverse of the density. Models which are at higher metallicity have larger radii and thus shallower outer envelopes, and hence larger
meridional currents. This favors mixing at higher metallicity. For single stars however, mass loss by stellar winds may strongly reduce
the angular momentum content and since meridional currents also scale with $\Omega$, mass losses may counteract the above effect, making
the mixing less strong at higher metallicity.
In close binary systems, the tidal locking may keep in the star an angular momentum larger than the one it would have if evolving without any tidal interaction. In that case, $\Omega$ is kept constant and everything being equal (same mass), the model with a higher metallicity will be more strongly mixed. 
Such an effect will be more marked in high mass star models which have a shallower envelope than in small initial mass stars.
\item Comparing the spin-up and the spin-down cases, we see that they are very similar given an orbital period. This is expected since the
synchronization timescale is short in both cases (although see the remark below) and the velocities to which the models converge
are similar since imposed by the orbital period. A careful comparison between these two cases show however a slight asymmetry between the spin-down and the spin-up cases: the spin-up cases are in general more mixed than the spin-down ones. These differences are mostly due to the fact that spin-up models starts with initial angular velocities that are slightly nearer from the orbital period than the spin-down models (see Tables A.1 and A.2).  Moreover, as explained in the Appendix B, in case of solid body rotation, the synchronization time is shorter in case of spin-up than in case of spin-down.
\item All the 60 M$_\odot$ models at Z=0.002 avoid a Roche lobe overflow during the MS phase (for the orbital periods considered here).
\end{itemize}

If we look at the panels for the two intermediate masses, 40 and 50 M$_\odot$, we obtain qualitative similar behaviors. We can just note that,
given an orbital period, higher the initial mass, stronger  the effects of the tidal forces are (let us remind here that in all the cases the companion has a mass equal to 2/3 the mass of the primary). 

In the right panel of Fig.~\ref{synt}, we indicate the models that show no Roche Lobe Overflow during the MS phase. We can note a few interesting points:
\begin{itemize}
\item Except in one case (the spin-down case for the 50 M$_\odot$ at Z=0.002), every time a RLOF is avoided during the MS phase,
the star has a homogeneous evolution.  On the other hand, not all models having a homogeneous evolution 
avoid a RLOF episode during the MS phase. Therefore the region of RLOF avoidance during the MS phase  is in general smaller than the region of homogeneous evolution. This comes from the two opposite conditions for homogeneous evolution and
RLOF avoidance:  RLOF avoidance is more difficult when the initial periods are short,  while homogeneous evolution is favored by short periods.
\item RLOF avoidance is favored for higher masses (more mixed) and lower metallicity (more compact) models.
\item Due to differences in the initial conditions for the spin-up and spin-down cases, as well as due to differences in the
synchronization timescales we have different situations in case of spin-down and spin-up. For this feature, initial conditions
have a significant impact. In general the spin-up cases more often avoids a RLOF episode during the MS phase.
\end{itemize}

We discussed above the impact on the efficiency of the mixing 
of changing the mass ratio, keeping the mass of the primary and the initial orbital period the same.
We indicated that a higher mass ratio favors shorter synchronization time and thus stronger mixing. This effect favors RLOF avoidance.
But one has also to account for the variation of the Roche limit with the mass ratio.
For a given mass of the primary and a given initial orbital period, the Roche limit
is larger in systems with larger mass ratios\footnote{This comes from the fact that to keep the same orbital period in systems with
more massive components, the distance between the two components needs to be larger.}.
This implies  that a higher mass ratio favors RLOF avoidance.


\section{Consequences of tidal interactions}

As explained in the previous section, tidal interactions in sufficiently close systems trigger strong mixing. 
In the present models, this strong mixing is induced by the high rotation imposed by the tidal coupling. 

We may wonder whether these binary models have an evolution equivalent
to a single star starting its evolution on the ZAMS with a rotation equal to the one obtained after
synchronization in the close binary system? The answer is no especially at high metallicity.
The main difference between these two cases is that the single star will lose a lot of angular momentum by stellar winds
while the component in the close binary system will lose mass by stellar winds but not angular momentum, since
the angular momentum lost by the winds will be replenished at the expense of the orbital angular momentum by
the tidal interactions. Therefore in these systems, the close binary interaction is not only important to
provide the star with a given initial rotation but also for maintaining this rotation all along the evolution until
Roche lobe episode occurs. 

The strong mixing may induce a homogeneous evolution and thus, as was already found by \citet{deMink09a},
may make the primary to avoid Roche lobe overflow episode. From the present results, we can deduce the 
following conditions favoring such a scenario: first obviously the orbital period should not be too small. For a given 
binary system, indeed, smaller the orbital period, more difficult it will be to avoid Roche lobe overflow even
when stars are strongly mixed. Second, conditions are more favorable in more massive systems, which are
those in which mixing is the strongest. We see for instance that no one of the binary systems containing a 30 M$_\odot$
avoids Roche lobe overflow during the MS phase, whatever the metallicity considered, while for instance
many cases of the systems with a 60 M$_\odot$ models avoids such an overflow during the MS phase.
Finally, we see that a lower metallicity constitutes also a favorable condition for avoiding Roche lobe overflow. 
Typically all the considered binary system with a 60 M$_\odot$ avoids Roche lobe overflow during the MS phase at Z=0.002,
while at Z=0.014, only the case with an orbital period equal to 1.8 days avoids it. 
Note that this last point may be surprising in view of the conclusion reached in the previous section where it was
indicated that models in close binary are more mixed at higher metallicities. We could have expected that if they are
more mixed, they keep more compact and thus would more easily avoids the Roche lobe overflow. But things are not so simple
indeed. Actually stars are more mixed at higher metallicities because they have a more extended envelope and a more
extended envelope makes the reaching of the Roche limit more easy! At low metallicities, stars remain more compact
and this favors the avoidance of the Roche lobe overflow.


We saw in the present work, that tidal interactions can provide to the primary in close binary systems sufficient angular momentum
for triggering a homogeneous evolution. This implies that the radiative envelope is rich in CNO-processed material at early times along the MS phase. Some mass will be lost by the stellar winds and some mass can be lost at the moment of the mass transfer
on the secondary in case the mass transfer is non-conservative. This material, lost by the system, if used to form new stars, will produce stars with some signature of CNO processing.
This might be a possible channel to explain second generation stars in globular clusters. 

Close binaries have already been invoked in this context by \citet{deMink09b}. These authors propose 
that the secondary is accelerated to the critical velocity through mass accretion from the primary. When the critical limit is reached,
the mass lost by the primary can no longer be accreted by the secondary and is therefore lost in the interstellar medium.
At that point, the primary may have lost sufficient mass for layers having been processed at least partially by the CNO burning
to be uncovered. This mass constitutes material that
can be used to form new stars that would show surface abundances with signatures of CNO processing. So in that scenario, the CNO processed matter is extracted from the primary because previous mass losses have uncovered H-burning regions and the mass is lost
because the secondary has reached the critical velocity. 
In the scenario we propose, CNO processed material is obtained very soon during the evolution due to the mixing triggered by tidal interactions. The mass at the moment of the mass transfer
can be lost because of the same reason invoked by de Mink (the reaching of the critical limit by the secondary) or other effects. Typically, the secondary, if massive enough,  loses some mass by stellar winds. This may
prevent it to accrete part or all the material lost from the primary. We shall discuss  the impact of such models in the frame of the origin of the anticorrelations in globular clusters in a forthcoming paper.


Could such an evolution be a plausible scenario for long soft Gamma Ray Bursts? At the moment, we cannot provide an answer to that question since it requires the computation of more advanced stages of the evolution of the systems. However, it is interesting that tidal interactions can 
trigger homogeneous evolution of the primary, makes it to enter the WR phase at an early stage in its evolution, while keeping a high angular momentum content. 
Of course the story is not finished and it may be that there will be still some mass transfer when the secondary would evolve in redder parts of the HRD. In cases however where no further mass transfer occurs or, if it occurs, that most of the mass would be lost by the system, then the present scenario could be an interesting one for making fast rotating black holes from objects with no hydrogen-rich envelope. Interestingly, such a scenario 
might explain the existence of long soft gamma ray burst even at high metallicity. Single fast rotating models with strong internal coupling lose a lot of angular momentum by stellar winds at high metallicity 
making the conditions for obtaining simultaneously a fast rotating black hole from a progenitor having lost its H-rich envelope very difficult.
Here, as we saw, tidal interactions can do two things that are favorable for obtaining these conditions, it produces mixing and, even when mass is lost, it replenishes the star
in spin angular momentum.

How can the present models be tested by observations? 
Present models can be checked through the observations of the surface abundances of short period binary systems before any mass transfer has occurred. These systems will be most likely synchronized in view of the very short timescale for this process to occur. The determination of the orbital period will determine also the axial rotation of the two components. If some indications are obtained on the initial mass, metallicity of the system, then the observed surface abundances can be compared to the predictions of the present computations. 
Note that,  whatever was the initial rotation of the primary, the surface abundances after synchronization will be the same. So the initial rotation of the components is not a critical parameter which is indeed a nice feature since this parameter cannot be estimated
in a model independent way. Thus these close binary models represent very interesting targets to check rotational mixing \citep[see also][]{deMink09a}.

Another  prediction of the present models is that the primary of all
close binary systems with an orbital period equal or  inferior to 1.4 days, with a primary more massive than 40 M$_\odot$, a companion having a mass equal to 2/3 of the primary, for metallicities
$Z$ equal or larger than 0.002 will follow a homogeneous evolution. Stars that follow a homogeneous evolution are overluminous for their mass.
All these models at $Z=0.002$ avoid a Roche Overflow event during the MS phase.

Close binary evolution may produce very fast rotating Wolf-Rayet (WR) stars as can seen from Tables A.1 and A.2.
Indeed, equatorial velocities between 160 and 360 km s$^{-1}$ can be obtained through the binary channel. The corresponding
single star models entering a WR phase have surface velocities at that stage between typically  20 and 80 km s$^{-1}$.
\citet{Grafener12} deduced rotational properties inferred from photometric variability for 
six WR stars (see their table 3 and references therein). Three of them present surface velocities between 110 and 230 km s$^{-1}$.
Such high surface velocities would be more in agreement with a close binary evolution.

\section{Conclusions and perspectives}


In this work, we studied the evolution of single and binary stellar models with a strong internal coupling. 

For single stars, we showed that higher the initial mass, the initial rotation, lower the metallicity, stronger the mixing efficiency.
This is exactly the same as in models with a moderate coupling, although the physics for the mixing
between these two kinds of models is very different.
In mildly coupled models, mixing is triggered by
the gradients of $\Omega$, while in strongly coupled models, it is driven by $\Omega$.
We obtain also that stars with a strong internal coupling, all other ingredients kept the same, 
are more strongly mixed than models with a moderate coupling.

When a companion is present, we checked that,  as is well known, the synchronization time is very short. 
So that the main effect of the companion will be to lock the rotation around the orbital rotation. 
The evolution after the synchronization process depends on the angular momentum at the synchronization time. The higher this angular momentum content is, the higher the degree of mixing. There is little dependence of the results on the initial velocity of the star.
The angular momentum after synchronization is of course the highest in the systems with the shortest orbital periods.


Large differences between single and close binary models with initial masses above about 30 M$_\odot$ having similar initial rotation
occur at large metallicities. 
The main difference comes from the fact that the single star model is spun down by losses of angular momentum by the stellar winds,
while the angular momentum lost by the star in the close binary system is replenished by the tidal interactions. In that case, mass losses
induce an acceleration of the star once synchronization is achieved.

We note also that, in contrast with the single star models, higher the metallicity, more efficient the mixing in the close binary models.

We have obtained some indications for the limiting orbital period below which our models will follow a homogeneous evolution.
The conditions for homogeneous evolution are more restricted at low metallicity for the reason just mentioned above
(less efficient mixing in low metallicity environments).

Quite generally, when no Roche Lobe Overflow occurs during the MS phases, the star has a homogeneous evolution.
Not all models evolving homogeneously avoid a RLOF episode. 
RLOF avoidance is favored in high mass star models at low metallicity.

How can we check these models through observations?
Ideally, one would like to observe close eclipsing binaries for which it is possible to obtain
the masses of the components and from spectroscopy their position in the HR diagram as well as some indications
of their surface abundances. These systems should be observed at a stage before any mass transfer. They would be
likely synchronized since synchronization time is too short for allowing the observation of a system before it is synchronized.

From an analysis of these data, we could check the following points: first, we can see whether these stars
present surface enrichments. We see that for short enough periods, surface enrichments are predicted for many
models, with or without a magnetic field \footnote{Note that stars following a homogeneous evolution 
show a  carbon to hydrogen ratio at the surface that  decreases in a first phase and then increases.
This non-monotonic behavior comes from the fact that, in these nearly homogeneous models, both carbon and hydrogen vary in an important manner.
Since the CN equilibrium is reached very rapidly, the first phase corresponds to the decrease of carbon while the abundance of hydrogen changes little. Thus the C/H ratio decreases. the second increasing part of the curve comes from the fact that carbon remains more or less constant (it has reached the CNO equilibrium value), while the abundance of hydrogen decreases. This makes this ratio to increase with time.
}. Second we can check whether stars are overluminous for their initial mass, indicating that indeed
they are strongly mixed. Third we can deduce their surface velocities. This last point is very important to
discriminate between two classes of models: actually the present models with strong internal coupling
predict strong mixing only for models with a high surface velocities (this is because the triggering of the mixing
is $\Omega$ and not the gradient of $\Omega$). On the other hand, models with mild internal mixing mainly driven by shear
can produce models which are strongly mixed and show a low surface velocity \citep{Song13}.
Note that these characteristics can also be obtained in single star evolution models when braking occurs by
a magnetic coupling between the surface of the star and the stellar wind \citep{Meynet11}.

Finally let us mention that the present evolutionary scenario might be interesting to consider in the frame of the
origin of the anticorrelation in globular clusters and for explaining the progenitors of some long soft Gamma Ray Bursts
especially those which might appear in metal-rich regions. Close binary evolution seems to be needed to explain
the existence of fast rotating Wolf-Rayet stars.

\begin{acknowledgements}
The authors thank the anonymous referee who helped through his very constructive remarks to improve this paper.
This work was sponsored by the National Natural Science Foundation of China (Grant No. 11463002) and the Key Laboratory for the Structure and Evolution of Celestial Objects, Chinese Academy of Sciences (Grant No. OP201405).
This work was supported by the Swiss National Science Foundation (project number 200020-146401).
\end{acknowledgements}

\appendix
\section{Some properties of the stellar models}

Some properties of the single and binary  models computed in this work are presented in Tables A.1 and A.2: The first four columns specify the model considered providing information about its status as a single or a component in a binary system
with a given orbital period, the ratio of the initial surface angular velocity to the critical velocity on the ZAMS, the ratio of the initial surface angular velocity
to the orbital one, as well as the initial equatorial rotation velocity. The five following columns indicate respectively the different times defined at the beginning of Section~4.
The last three columns give some properties of the last computed model: $X_c({\rm end})$ is the mass fraction of hydrogen at the center, $\upsilon_{\rm eq}({\rm end})$ is the surface equatorial velocity and N/C(end) 
the ratio at the surface of the mass fractions of nitrogen to carbon normalized to the initial ratio. The initial ratio is the same for the three metallicities and is equal to 0.29 in mass fraction.

\begin{table*}
\scriptsize{ 
\caption{Some properties of the single and binary models for initial masses equal to 30 M$_\odot$ and 40 M$_\odot$. In case the model is in a binary, the companion has an initial mass of respectively 20 and 26.7  M$_\odot$ .
The symbol S is for single star models and the symbol BN.N is for close binaries with an orbital period equal to N.N days. For each metallicity and initial mass, eight models have been computed, four
with a high initial rotation and four with a moderate initial rotation. When in binaries, the fast rotating cases correspond to tidally spin-down cases, while the moderately rotating ones correspond to
spin-up cases. A small line in the column $t_{\rm WR}$  indicates that the star does not become a WR star before any RLOF or during the MS phase.} \label{tab1} \centering
\begin{tabular}{cccc|cccccccc}
\hline\hline
Model & ${\Omega_{\rm ini}\over \Omega_{\rm crit}}$ & ${\Omega_{\rm ini}\over \Omega_{\rm orb}}$ &   $\upsilon_{\rm ini,eq}$ & $t_{\rm sync}$ & $t_{N/C}$                          & $t_{\rm hom}$ & $t_{\rm WR}$  & $t({\rm end})$ & $X_c({\rm end})$ & $\upsilon_{\rm eq}({\rm end})$      & N/C({\rm end}) \\
           &  &                                                                                  & km s$^{-1}$                      & $10^6$ yr         &  $10^6$ yr                         & $10^6$ yr       & $10^6$ yr         & $10^6$ yr       &                                 &km s$^{-1}$                                        &                             \\
\hline
\multicolumn{12}{c}{} \\
\multicolumn{12}{c}{\underline{30 (+20) M$_\odot$} }\\
\multicolumn{12}{c}{} \\
\multicolumn{12}{c}{$Z=0.002$} \\
\multicolumn{12}{c}{} \\
S                 & 0.66  &   --                               &   391& --      & 0.81    &  1.76  &    7.12     & 8.24   &   0.035  &  79    & 307 \\
B1.4            & 0.66 &   1.772                       &   394& 0.34 & 0.45    &  1.35  &      --        & 6.17   &   0.24     &  292  & 272 \\
B1.6            & 0.54 &   2.025                       &   311& 0.54 & 0.40    &  1.25  &      --        & 5.75   &   0.25     &  267  & 131 \\
B1.8            & 0.66 &   2.279                       &   394& 0.81 & 0.38    &  1.21  &      --        & 5.64   &   0.22     &  242  &  90   \\
\multicolumn{12}{c}{} \\
S                  & 0.30 &   --                                 &   163& --      & 2.54    &  1.08  &      --        & 6.38   &   0.041   &  142   & 17  \\
B1.4            & 0.30 &   0.769                          &   167& 0.17 & 0.93    &  1.23  &      --        & 6.02   &   0.26     &  292  & 186 \\
B1.6            & 0.30 &   0.879                          &   164& 0.17 & 1.22    &  1.12  &      --        & 5.54   &   0.26     &  267  &   69 \\
B1.8           &  0.30 &   0.989                          &   164& 0.00 & 1.41    &  1.09  &      --        & 5.45   &   0.24     &  242  &   41  \\
\multicolumn{12}{c}{} \\
\multicolumn{12}{c}{$Z=0.007$} \\
\multicolumn{12}{c}{} \\
S                  & 0.66 &   --                                 &   374& --      & 0.83    &  1.75  &      --       & 8.02   &   0.035  &  58     & 369 \\
B1.4            & 0.66 &   1.551                       &   377& 0.23 & 0.59    &  1.48  &     6.15   & 6.96   &   0.19     &  285  & 345 \\
B1.6            & 0.66 &   1.773                       &   377& 0.36 & 0.60    &  1.32  &      --        & 5.59   &   0.30     &  270  & 131 \\
B1.8            & 0.66 &   1.995                       &   377& 0.54 & 0.58    &  1.25  &      --        & 5.37   &   0.29     &  248  &   76  \\
\multicolumn{12}{c}{} \\
S                  & 0.30 &   --                                 &   163& --      & 2.54    &  1.08  &      --        & 6.38   &   0.040   &   58   & 10   \\
B1.4            & 0.30 &   0.769                          &   167& 0.17 & 0.93    &  1.23  &      --        & 6.88   &   0.20     &  286  & 341 \\
B1.6            & 0.30 &   0.879                          &   164& 0.17 & 1.22    &  1.12  &      --        & 5.47   &   0.31     &  270  &   86  \\
B1.8            & 0.30 &   0.989                          &   164& 0.00 & 1.41    &  1.09  &      --        & 5.18   &   0.30     &  249  &   41  \\
\multicolumn{12}{c}{} \\
\multicolumn{12}{c}{$Z=0.014$} \\
\multicolumn{12}{c}{} \\
S                  & 0.66 &   --                                 &   362& --      & 0.92    &  1.55  &      --       & 7.62   &   0.041   &  23     & 369 \\
B1.4            & 0.66 &   1.411                       &   365& 0.18 & 0.64    &  1.53  &     6.20   & 6.73   &   0.24     &  281  & 345 \\
B1.6            & 0.66 &   1.613                       &   365& 0.29 & 0.69    &  1.33  &      --        & 5.12   &   0.36     &  270  & 131 \\
B1.8            & 0.66 &   1.815                       &   365& 0.43 & 0.71    &  1.25  &      --        & 4.86   &   0.35     &  250  &   76  \\
\multicolumn{12}{c}{} \\
S                  & 0.30 &   --                                 &   151& --      & 3.44    &  1.01  &      --        & 6.24   &   0.056   &   31   &   9   \\
B1.4            & 0.30 &   0.622                          &   151& 0.11 & 0.61    &  1.48  &      --        & 6.72   &   0.24     &  281  & 379 \\
B1.6            & 0.30 &   0.711                          &   151& 0.38 & 0.97    &  1.22  &      --        & 5.03   &   0.37     &  270  &   93  \\
B1.8            & 0.30 &   0.800                          &   151& 0.45 & 1.31    &  1.10  &      --        & 4.71   &   0.36     &  251  &   38  \\
\multicolumn{12}{c}{} \\
\hline
\multicolumn{12}{c}{} \\
\multicolumn{12}{c}{\underline{40 (+26.7) M$_\odot$} }\\
\multicolumn{12}{c}{} \\
\multicolumn{12}{c}{$Z=0.002$} \\
\multicolumn{12}{c}{} \\
S                  & 0.66 &   --                                 &   416& --      & 0.51    &  1.88  &   5.28     & 6.24   &   0.035    &  57    & 286 \\
B1.4            & 0.66 &   1.607                         &   419& 0.16 & 0.23    &  1.35  &   5.02     & 6.00   &   0.036    &  207  & 286 \\
B1.6            & 0.66 &   1.837                         &   419& 0.24 & 0.20    &  1.12  &      --        & 5.37   &   0.15     &  302  & 362 \\
B1.8            & 0.66 &   2.066                         &   419& 0.36 & 0.18    &  1.02  &      --        & 4.84   &   0.20     &  280  & 224   \\
\multicolumn{12}{c}{} \\
S                  & 0.30 &   --                                 &   173& --      & 1.54    &  0.84  &      --        & 5.10   &   0.041   &  129   & 23  \\
B1.4            & 0.30 &   0.704                          &   174& 0.07 & 0.33    &  1.25  &    5.0      & 5.98   &   0.040    &  205  & 286 \\
B1.6            & 0.30 &   0.804                          &   174& 0.13 & 0.55    &  0.99  &      --        & 5.25   &   0.16     &  302  &  303 \\
B1.8            & 0.30 &   0.905                          &   174& 0.00 & 0.70    &  0.90  &      --        & 4.70   &   0.22     &  280  &  114  \\
\multicolumn{12}{c}{} \\
\multicolumn{12}{c}{$Z=0.007$} \\
\multicolumn{12}{c}{} \\
S                  & 0.66 &   --                                 &   397& --      & 0.51    &  1.57  &     6.09   & 6.09   &   0.035  &  24     & 344 \\
B1.4            & 0.66 &   1.401                         &   401& 0.08 & 0.30    &  5.03  &     4.89   & 5.03   &   0.25     &  272  & 344 \\
B1.6            & 0.66 &   1.601                         &   401& 0.16 & 0.30    &  1.29  &     5.13    & 5.29   &   0.19     &  245  & 341 \\
B1.8            & 0.66 &   1.802                         &   401& 0.24 & 0.27    &  1.10  &      --        & 4.69   &   0.25     &  277  &  248 \\
\multicolumn{12}{c}{} \\
S                  & 0.30 &   --                                 &   166& --      & 1.71    &  0.80  &      --        & 5.10   &   0.040     &   29   & 19   \\
B1.4            & 0.30 &   0.617                          &   166& 0.06 & 0.24    & 5.28  &      4.89   & 6.14   &   0.041     &  208  & 314 \\
B1.6            & 0.30 &   0.705                          &   166& 0.10 & 0.39    &  1.20  &      5.14  & 6.02   &   0.041     &  190  &  314  \\
B1.8            & 0.30 &   0.793                          &   166& 0.20 & 0.60    &  0.97  &      --        & 4.62   &   0.25      &  277 &   274 \\
\multicolumn{12}{c}{} \\
\multicolumn{12}{c}{$Z=0.014$} \\
\multicolumn{12}{c}{} \\
S                  & 0.69 &   --                                 &   401& --      & 0.55    &  1.31  &      --       & 5.21   &   0.18   &  33     & 310 \\
B1.4            & 0.66 &   1.268                         &   387& 0.06 & 0.32    &  4.16  &     --        & 4.16   &   0.37     &  318  & 386 \\
B1.6            & 0.66 &   1.450                         &   387& 0.13 & 0.34    &  1.40  &    4.92    & 4.92   &   0.26     &  278  & 369 \\
B1.8            & 0.66 &   1.631                         &   387& 0.19 & 0.34    &  1.12  &      --        & 4.32   &   0.31     &  275  & 286  \\
\multicolumn{12}{c}{} \\
S                  & 0.30 &   --                                 &   160& --      & 1.93    &  0.77  &      --        & 4.81   &   0.095     &   17   &   17   \\
B1.4            & 0.30 &   0.540                         &   161& 0.05 & 0.20    &  4.16  &      --        & 4.16   &   0.37       &  318  & 386 \\
B1.6            & 0.30 &   0.640                          &   161& 0.08 & 0.32    &  1.36  &    4.91    & 6.14   &   0.040     &  160  & 338  \\
B1.8            & 0.30 &   0.720                          &   161& 0.24 & 0.50    &  1.03  &      --        & 4.29   &   0.31      &   275 &  248 \\
\multicolumn{12}{c}{} \\
\hline
\hline

\end{tabular}
}
\end{table*}

\begin{table*}
\scriptsize{ 
\caption{Some properties of the single and binary models for initial masses equal to 50 M$_\odot$ and 60 M$_\odot$. In case the model is in a binary, the companion has an initial mass of respectively 33.3 and 40  M$_\odot$. Same comments as in the caption of Table 1.} \label{tab1} \centering
\begin{tabular}{cccc|cccccccc}
\hline\hline
Model & ${\Omega_{\rm ini}\over \Omega_{\rm crit}}$  & ${\Omega_{\rm ini}\over \Omega_{\rm orb}}$ &   $\upsilon_{\rm ini,eq}$ & $t_{\rm sync}$ & $t_{N/C}$                          & $t_{\rm hom}$ & $t_{\rm WR}$  & $t({\rm end})$ & $X_c({\rm end})$ & $\upsilon_{\rm eq}({\rm end})$      & N/C({\rm end}) \\
            & &                                                                                  & km s$^{-1}$                      & $10^6$ yr         &  $10^6$ yr                         & $10^6$ yr       & $10^6$ yr         & $10^6$ yr       &                                 &km s$^{-1}$                                        &                             \\
\hline
\multicolumn{12}{c}{} \\
\multicolumn{12}{c}{\underline{50 (+33.3) M$_\odot$} }\\
\multicolumn{12}{c}{} \\
\multicolumn{12}{c}{$Z=0.002$} \\
\multicolumn{12}{c}{} \\
S          &    0.66    &   --                               &   436& --      & 0.35    &  2.14  &    4.32     & 5.19   &   0.035     &  46    & 286 \\
B1.4    &    0.66    &   1.490                       &   440& 0.07 & 0.16    &  4.71  &    3.96     & 5.01   &   0.035     &  219  & 266 \\
B1.6     &   0.66    &   1.702                       &   440& 0.12 & 0.14    &  1.25  &    4.10     & 4.98   &   0.036     &  191  & 266 \\
B1.8     &   0.66     &   1.915                       &   440& 0.18 & 0.13    &  1.00  &    --          & 4.78   &   0.08        &  299  & 183  \\
\multicolumn{12}{c}{} \\
S            &  0.30    &   --                                 &   182& --      & 1.07    &  0.69  &      --        & 4.40   &   0.040   &  98.7   & 31  \\
B1.4      &  0.30    &   0.654                          &   183& 0.05 & 0.17    & 4.71  &      --        & 4.98   &   0.040  &  222& 266 \\
B1.6       & 0.30    &   0.748                          &   183& 0.07 & 0.17    &  1.11  &     4.1    &  4.97  &   0.040  &  194 &  266 \\
B1.8       & 0.30     &   0.841                          &   183& 0.10 & 0.39    &  0.88  &      --        & 4.62   &   0.12     & 307  & 286  \\
\multicolumn{12}{c}{} \\
\multicolumn{12}{c}{$Z=0.007$} \\
\multicolumn{12}{c}{} \\
S            &  0.66    &   --                              &   416& --      & 0.37    &  1.51  &     5.01   & 5.06   &   0.035  &  35     & 34 \\
B1.4       &  0.66   &   1.294                       &   420& 0.04 & 0.19    & 4.05  &     3.97   & 4.05   &   0.27     &  306  & 345 \\
B1.6      &  0.66    &   1.479                       &   420& 0.09 & 0.19    & 4.26  &     3.98    & 4.26   &   0.22     &  256  & 334 \\
B1.8      &  0.66    &   1.664                       &   420& 0.12 & 0.19    & 1.23  &     4.13    & 4.23   &   0.22     &  248  & 334  \\
\multicolumn{12}{c}{} \\
S            &  0.30    &   --                                 &   174& --      & 1.17    &  0.67  &      --        & 4.37   &   0.041   &   19   & 31   \\
B1.4      &  0.30    &   0.569                         &   174& 0.04 & 0.13    &  4.64  &     3.9     & 5.18    &   0.041  &  205  & 303 \\
B1.6       & 0.30    &   0.650                          &   174& 0.05 & 0.19    & 4.50  &     3.98    & 5.12   &   0.041   & 178  & 307  \\
B1.8       & 0.30    &   0.732                          &   174& 0.13 & 0.29    &  1.14  &    4.14     & 5.04   &   0.041    &158  & 307  \\
\multicolumn{12}{c}{} \\
\multicolumn{12}{c}{$Z=0.014$} \\
\multicolumn{12}{c}{} \\
S            &   0.66   &   --                              &   403& --      & 0.39    &  1.15  &      --       & 4.84   &   0.035   &  14     & 331 \\
B1.4      &   0.57   &   1.138                       &   338& 0.00 & 0.21    &  1.63  &     --        & 1.63   &   0.58     &  356  & 34 \\
B1.6      &   0.57   &   1.301                       &   338& 0.04 & 0.23    &  3.97  &     3.94    &3.97   &   0.27     &  291  & 355 \\
B1.8      &   0.57   &   1.404                       &   338& 0.07 & 0.26    &  1.44  &     3.98    & 4.06   &   0.28     &  235  & 355  \\
\multicolumn{12}{c}{} \\
S            &    0.30  &   --                                 &   168& --      & 1.29    &  0.64  &      --        & 4.11       &   0.091    &   12   &   28   \\
B1.4     &     0.30  &   0.517                          &   169& 0.04 & 0.11    & 1.63  &      --            & 1.63   &   0.57       &  356  & 383 \\
B1.6      &    0.30  &   0.590                          &   169& 0.04 & 0.16    &  4.38  &    3.94        & 5.24   &   0.041     &  153  & 314  \\
B1.8      &    0.30  &   0.664                          &   169& 0.10 & 0.24    &  1.41  &    3.98     & 5.14      &   0.041     &  134  & 334  \\
\multicolumn{12}{c}{} \\
\hline
\multicolumn{12}{c}{} \\
\multicolumn{12}{c}{\underline{60 (+40) M$_\odot$} }\\
\multicolumn{12}{c}{} \\
\multicolumn{12}{c}{$Z=0.002$} \\
\multicolumn{12}{c}{} \\
S             &   0.66  &   --                                 &   454& --      & 0.28    &  2.17  &   3.74     & 4.54   &   0.035    &  41    & 266 \\
B1.4       &   0.67  &   1.401                         &   460& 0.04 & 0.14    &  4.24  &   3.41     & 4.40   &   0.035    &  234  & 266 \\
B1.6        &  0.67  &   1.601                         &   426& 0.08 & 0.12    &  4.13  &   3.41      &4.37   &   0.035   &   200 &  266 \\
B1.8       &   0.67  &   1.801                         &   394& 0.11 & 0.11    &  1.20  &   3.57      & 4.36   &   0.036    & 176  & 266   \\
\multicolumn{12}{c}{} \\
S              &   0.30  &   --                                 &   190& --      & 0.80    &  0.61  &      --        & 3.90   &   0.058   &    78   &    38  \\
B1.4          &  0.30 &   0.608                          &   212& 0.03 & 0.10    & 4.23  &   3.40      & 4.38   &   0.040    &  237  & 266 \\
B1.6         &  0.30  &   0.696                          &   224& 0.05 & 0.15    & 4.12  &   3.41      & 4.36   &   0.040   &   203 &  266 \\
B1.8        &   0.30 &   0.783                          &   190& 0.08 & 0.23    & 1.04  &   3.62      & 4.34   &   0.040   &   180 &  266  \\
\multicolumn{12}{c}{} \\
\multicolumn{12}{c}{$Z=0.007$} \\
\multicolumn{12}{c}{} \\
S              &  0.66  &   --                                 &   432& --      & 0.29    &  1.49  &     4.34   & 4.43   &   0.035    &  32     & 326 \\
B1.4        &  0.66  &   1.211                         &   437& 0.03 & 0.13    &  3.38  &      --        & 3.38   &   0.30      &  365  & 335\\
B1.6        &  0.66  &   1.384                         &   437& 0.06 & 0.14    &  4.10  &    3.42    & 4.56  &   0.035      &  181  & 305 \\
B1.8        &  0.66  &   1.557                         &   437& 0.08 & 0.14    &  3.96  &    3.43     & 4.51   &   0.036     &  158  &  305 \\
\multicolumn{12}{c}{} \\
S            &    0.30  &   --                                 &   181& --      & 0.86    &  0.60  &      --        & 3.89   &   0.048     &   16   & 13   \\
B1.4      &   0.30   &   0.534                          &   181& 0.03 & 0.08    & 4.23  &      3.42   & 4.59   &   0.040     &  211  & 89 \\
B1.6        & 0.30   &   0.611                          &   181& 0.04 & 0.11    & 4.09  &      3.41  & 4.53  &   0.041     &  183 &    88  \\
B1.8         & 0.30  &   0.687                          &   181& 0.06 & 0.16    & 3.95  &      3.42   & 4.49   &   0.041    &  161 &   91 \\
\multicolumn{12}{c}{} \\
\multicolumn{12}{c}{$Z=0.014$} \\
\multicolumn{12}{c}{} \\
S                & 0.66  &   --                                 &   418& --      & 0.30    &  1.05  &     --       & 4.26   &   0.035      &  12     &   27 \\
B1.4          & 0.67 &   1.085                         &   420& 0.00 & 0.13    &  0.16  &     --        & 0.16   &   0.70        &  388  &   4.4 \\
B1.6         &  0.69 &   1.240                         &   440& 0.04 & 0.15    &  3.54  &    3.39    & 3.54   &   0.26        &  330  & 351 \\
B1.8         &  0.69 &   1.395                         &   408& 0.06 & 0.16    &  4.15  &    3.39     & 4.45   &   0.035     &  255  & 296  \\
\multicolumn{12}{c}{} \\
S              &   0.30 &   --                                 &   175& --      & 0.93    &  0.57  &      --        & 3.62   &   0.10        &   10   &   13   \\
B1.4        &    0.30&   0.483                         &    175& 0.03 & 0.07    &  0.17  &      --        & 0.17   &   0.70       &  382  &   4.1 \\
B1.6         &  0.30 &   0.552                          &   175& 0.03 & 0.09    & 3.54  &    3.35    &  3.64 &   0.26          &  330  &  5.3  \\
B1.8        &    0.30 &   0.620                          &   175& 0.05 & 0.13    & 4.17  &    3.39     & 4.43   &   0.041      &   292 &  86 \\
\multicolumn{12}{c}{} \\
\hline
\hline

\end{tabular}
}
\end{table*}  

\section{Tidal interaction in close binary systems: impact of two different transport mechanisms for the angular momentum
}

In  paper I \citep{Song13}, we studied the effects of tidal interaction in close binary systems consisting of a 15  and  a 10 M$_\odot$ star. We considered that the mechanisms for the transport of the angular momentum inside the star were the meridional currents and the shear instabilities. 
Here, we computed the models accounting for the Tayler-Spruit dynamo which is a much more efficient transport mechanism for the angular momentum
and which imposes, at every time, a distribution of the angular velocity inside the star very close to a solid body rotation. In order to
see how these different transport mechanisms affect various properties of the single and close binary stars, we compare the results obtained for
15 M$_\odot$ models computed with these two methods and starting with exactly the same initial conditions.

\subsection{Spin-down case} 

\begin{figure*}
   \centering
    \includegraphics[width=8.5cm]{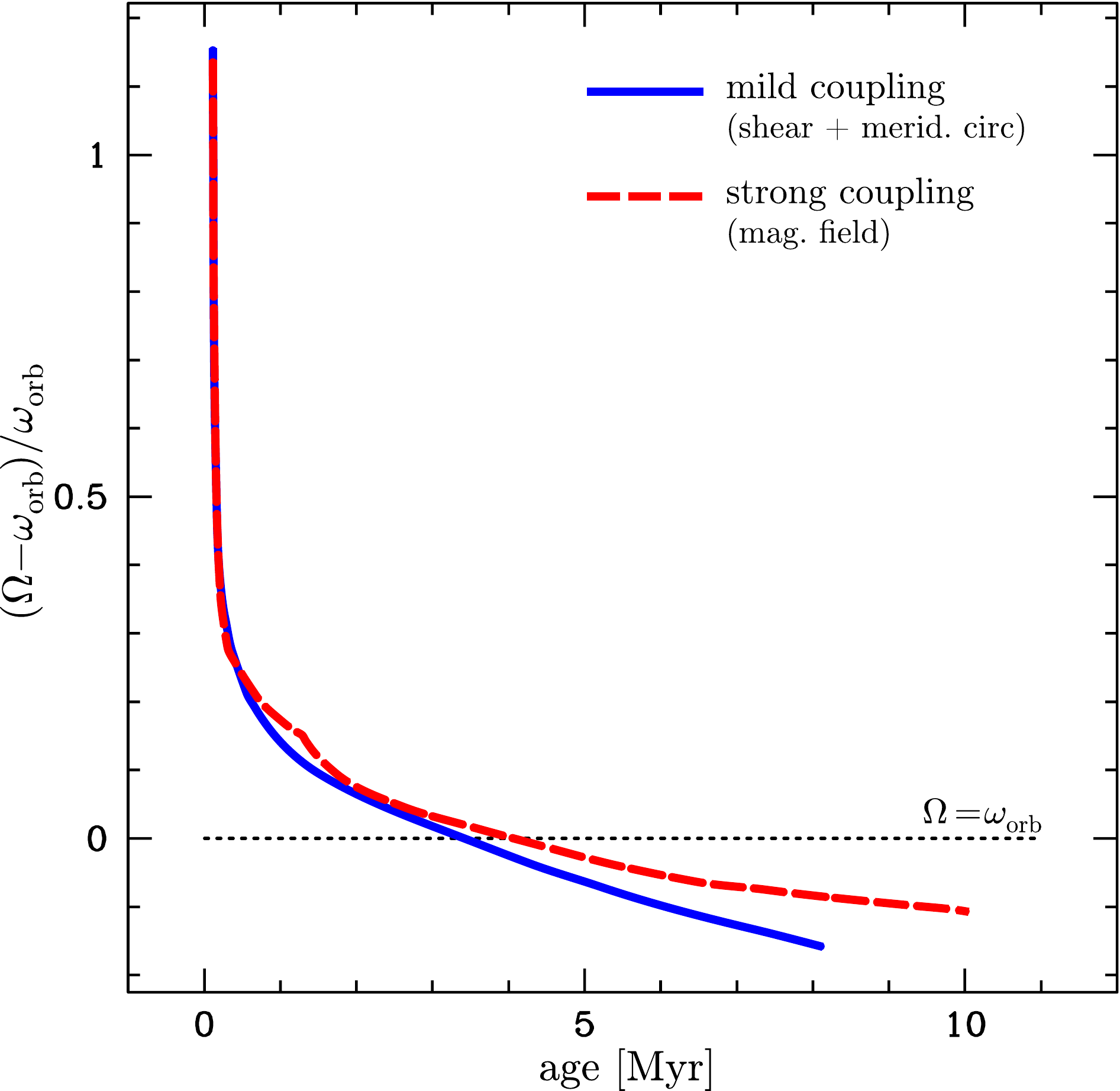}\hfill\hfill  \includegraphics[width=8.5cm]{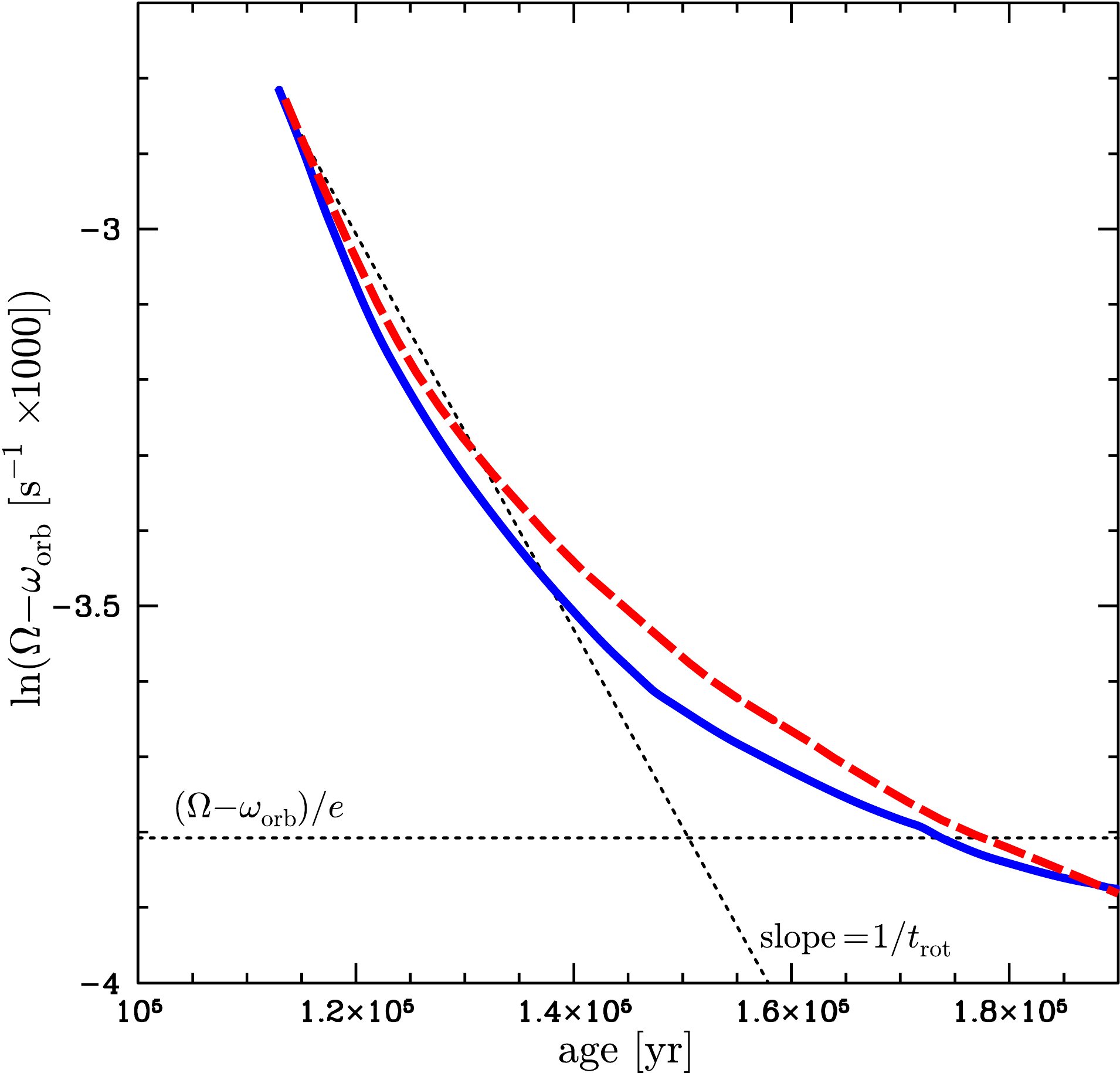}   
      \caption{{\it Left panel:} Evolution as a function of time of the surface angular velocity, $\Omega$, minus the orbital angular velocity, $\omega_{\rm orb}$, normalized to the orbital angular velocity for a 15 M$_\odot$ model at Z=0.014 beginning its evolution on the ZAMS with
      an initial rotation around 420 km s$^{-1}$ and an orbital period equal to 1.4 days. Its companion has 10 M$_\odot$.
      The continuous (blue) line corresponds to models computed with a mild coupling mediated by shear and meridional currents. The dashed (red) line corresponds to models with a strong coupling mediated by an internal magnetic field
      (solid body rotation). The horizontal dotted line has an ordinate equal to 0, {\it i.e.} its intersection with the other curves indicates when an equality between  $\Omega$ and $\omega_{\rm orb}$ is obtained.
      {\it Right panel:} Evolution as a function of time of $\Omega-\omega_{\rm orb}$ (natural logarithm) for the same models as in the left panel. The horizontal dotted line is at an ordinate equal to $(\Omega-\omega_{\rm orb})/e$.
      The other dotted line has a slope equal to $1/t_{\rm rot}$ as given by Eq.~(6) in Paper I.}
         \label{syncsd}
   \end{figure*}

\begin{figure*}
   \centering
    \includegraphics[width=8.5cm]{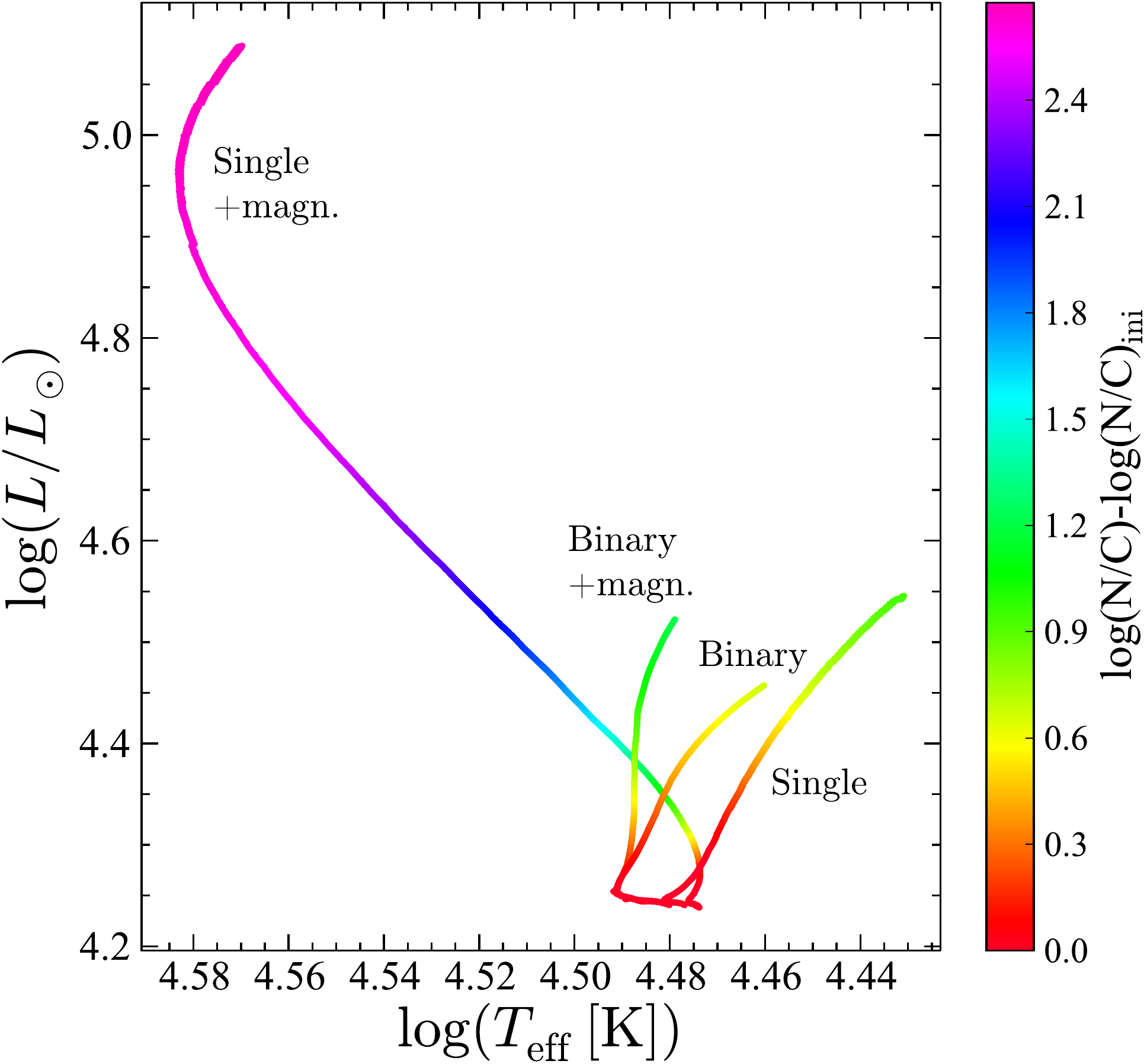}\hfill\hfill  \includegraphics[width=8.9cm]{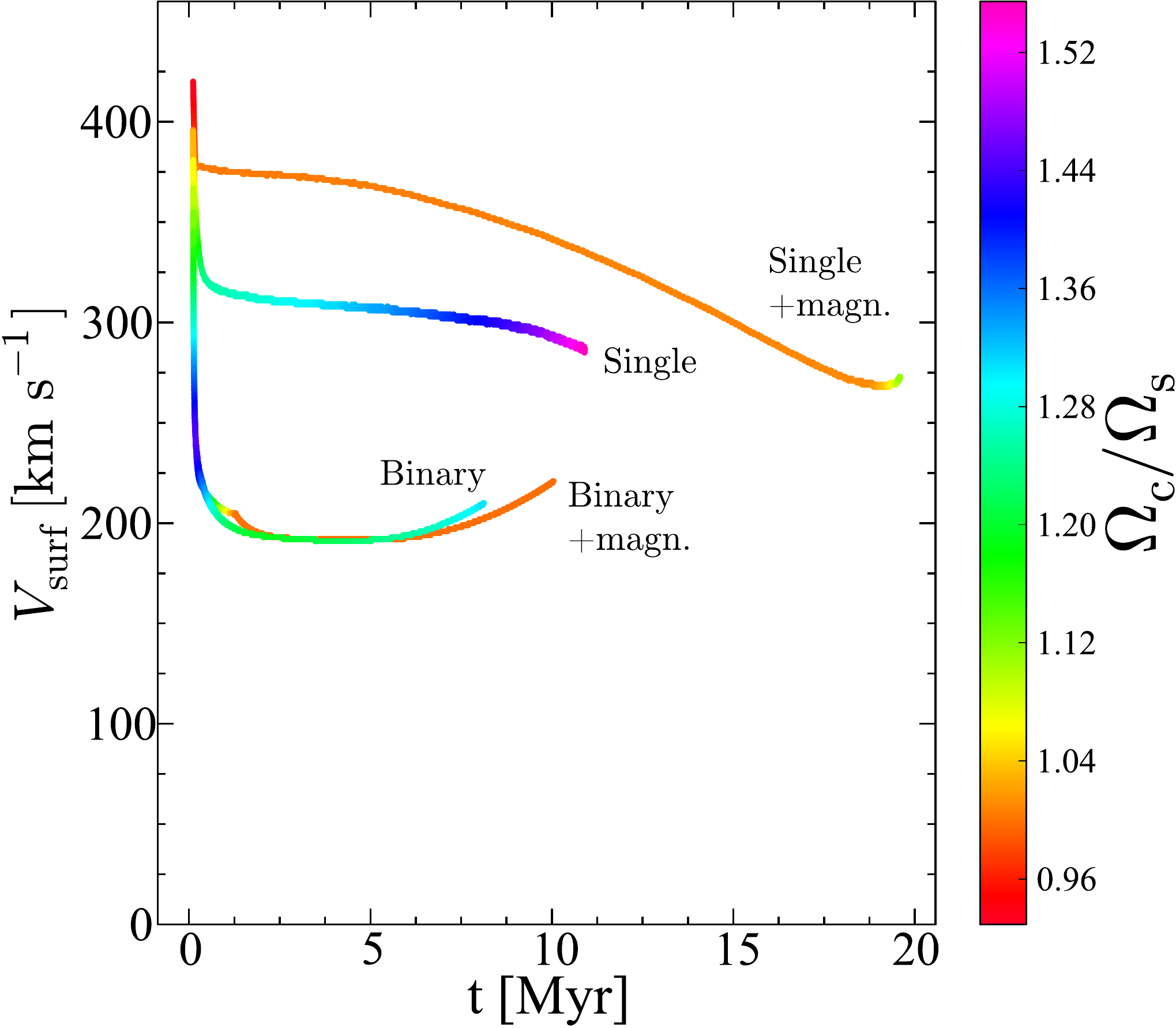}   
      \caption{{\it Left panel:} Evolutionary tracks in the HR diagram for 15 M$_\odot$ at Z=0.014 beginning their evolution on the ZAMS with
      initial rotation around 420 km s$^{-1}$. The track labeled {\it Single} is for an isolate star computed as in Paper I (transport of angular momentum by shear and
      meridional currents), the one labels with {\it  Binary} is for a binary with an initial orbital period of 1.4 days computed with the same transport mechanisms as in Paper I.
      The models labeled {\it Single+magn.} and {\it Binary+magn.} are for single and binary stars respectively with similar characteristics but computed with the Tayler-Spruit dynamo.
      The color scale on the right shows the evolution of abundance ratio N/C at the surface normalized to the initial value on the ZAMS.
      {\it Right panel:} Evolution of the surface equatorial velocity for the same models as in the left panel. The color scale on the right shows the evolution of  the ratio
      $\Omega_c/\Omega_s$.}
         \label{dhr15comp}
   \end{figure*}

Let us begin discussing the spin-down case for a 15 M$_\odot$ model having a 10 M$_\odot$ companion 
with  an initial orbital period of 1.4 days. In Fig.~\ref{syncsd}, the evolution of the difference between the actual surface angular velocity $\Omega$ and the orbital angular velocity $\omega_{\rm orb}$
for the cases without (continuous line) and with an internal magnetic field (dashed line) are shown.
Note that the case without an internal magnetic field corresponds to a model already studied in paper I  \citep{Song13}.
The left panel of the figure shows the evolution of the difference between the actual surface angular velocity and the orbital one normalized to the orbital one. The right panel
is a zoom on the very beginning of the slowing down process and show the difference between the actual velocity and the orbital one with no normalization.

We see that the evolution is quite similar in models with and without an internal magnetic field. This comes from the fact that at the beginning the two models present very similar characteristics (e.g. both are solid body rotating). Actually, we can see that both models enter the regime where they have a surface angular velocity differing from the orbital one by less than 20\% very rapidly.
This illustrates the well known fact that indeed synchronization is a very rapid process \citep[see e.g. the discussion in][]{deMink09a}.

How the actual synchronization timescale obtained by the numerical models compares to the values obtained from analytic formulae?
An interesting quantity is  $t_{\rm rot}$ (see for instance Eq.~6 in Paper I); $t_{\rm rot}$ corresponds to the time that is needed to decrease the quantity  $\Omega-\omega_{\rm orb}$ by a factor $e$ (on the left panel of Fig.~\ref{syncsd} this corresponds to a decrease of one dex in the ordinate scale). 

For evaluating $t_{\rm rot}$ from the analytical formula, we use the characteristics of the models on the ZAMS. The formula involve the moment of inertia of the region where the
tidal force applies.
Thus, this timescale will vary a lot depending on which portion of the star is considered to estimate the moment of inertia. In case the moment of inertia of the whole star
is considered, the time is obviously longer than when only the moment of inertia of the outer layer is considered.  
Plugging the moment of inertia of the whole star in the analytic formally implicitly assumes that any change of the angular momentum at the surface has instantaneously
an impact on the angular momentum in the core. This corresponds to the strongest coupling we can imagine and this implies the longest synchronization time
(the tides have to slow down the whole star and not only its outer layers. When transport by magnetic fields are assumed, we have a situation approaching that of an instantaneous redistribution). 
The case of considering the moment of inertia of only the outer layers would correspond to the other extreme case where there is no coupling between the core
or the region below these outer layers and these outer layers. In that case, synchronization is obtained more rapidly since a much smaller mass needs to be spun up
or spun down. These examples show that numerical models with some moderate internal coupling will be in-between these two cases.

Is it the situation obtained here by the numerical models? Yes it is. To see that let us consider 
the specific case of the 15 M$_\odot$ model at Z=0.014 in close binary system with a companion of 10 M$_\odot$ and with an orbital period of 1.4 days.
The value for $t_{\rm rot}$ is equal to 38100 years plugging the moment of inertia of the whole star in the analytic formula (Eq.~6 in Paper I). 
This timescale corresponds to about 0.3\% of the Main-Sequence lifetime of a non-rotating solar metallicity 15 M$_\odot$ model, whose Main-Sequence lifetime is about 11 My.

Let us now determine from the numerical model the time needed to decrease by a factor $e$ the quantity $\Omega-\omega_{\rm orb}$. The value of $\omega_{\rm orb}$ corresponding to a period of 1.4 days is equal to 0.052 s$^{-1}$. The horizontal dotted line on the right panel of Fig.~\ref{syncsd} shows the values $(\Omega-\omega_{\rm orb})/e$.
We see that the intersection between the curve  showing $\Omega-\omega_{\rm orb}$ (continuous curve) and the dotted line occurs at a time t=0.174 Myr. Now the start time of the model is not 0 but the time $t_0=0.114$ Myr because the model needs some time to
settle into a ZAMS structure. So the duration for decreasing $\Omega-\omega_{\rm orb}$ by a factor e is equal to 174000 - 114000 = 60000 years.  Actually this is  above the value deduced from the analytic formula by about a factor 2. 

This might be at first sight surprising when considering that the analytic formula should already be an upper limit, but the analytic formula is strictly valid only on a timescale during which $\Omega-\omega_{\rm orb}$ is not varying too much. Actually, as can be seen from Fig.~\ref{syncsd}, $\Omega-\omega_{\rm orb}$ decreases very fast (actually $\Omega$ is varying very fast since $\omega_{\rm orb}$ keeps a constant value) and thus $t_{\rm rot}$ is varying very fast too. Since $\Omega$ tends towards $\omega_{\rm orb}$, $t_{\rm rot}$ increases, so that the actual time to decrease by a factor e can actually be larger than the value obtained from ZAMS properties. The increase is however modest. Indeed, 
60000 years remains a very short value compared to the MS lifetime (less than 1\%).

On the right panel of Fig:~\ref{syncsd}, the slope at the beginning should be near the value given by $1/t_{\rm rot}$.
Interestingly, we can see that the slope at the ZAMS  of the continuous line (model with shear and meridional currents without magnetic field) is steeper than the slope of the dotted line, indicating that
indeed at the beginning, the synchronization time of the numerical model is shorter than the one obtained by the analytic formula, as expected.
We see also that the model with an internal magnetic field (dashed line) has an intermediate position between the
continuous and the dotted line also exactly as expected.  This short discussion therefore demonstrates that the present results are indeed quite consistent.

We can note that the time when $\Omega$ is equal to $\omega_{\rm orb}$ is around 3.5 Myr and 4 Myr for the non-magnetic and magnetic models respectively (see the intersections
of the horizontal dotted line with the curves on the left panel of Fig.~\ref{syncsd}). However
from an observational point of view, the star can be said synchronized much earlier when typically the difference between these two angular velocities is less than 10-20\%.
Indeed, uncertainties pertaining the measures of the rotation velocity and of the stellar radius  (to obtain $\Omega$ from $\upsilon_{\rm rot}$) will make extremely difficult
to identify the point when $\Omega$ is equal to $\omega_{\rm orb}$. Moreover, as can be seen from Fig.~\ref{syncsd}, the decrease of $\Omega$ is very rapid and important at the beginning
making the star to be quasi-synchronized in a very short time interval. We note also that after the point when $\Omega=\omega_{\rm orb}$, $\Omega$ becomes slightly inferior to $\omega_{\rm orb}$.
This comes from the fact that the surface velocity of the star is governed by various processes with different timescales: the tidal forces which have a relatively long timescale
when $\Omega$ is near $\omega_{\rm orb}$, the inflation of the envelope (nuclear timescale), the mass loss (the process becomes more important when evolution proceeds and has a rapid timescale), the internal transport mechanisms (which are rapid in that case).
The global effect of all these processes on the surface angular velocity results from their interplay. It is natural that when $\Omega$ is near $\omega_{\rm orb}$, the tidal forces likely do no longer dominate, at least until
the differences between those two quantities are not too large, and thus $\Omega$ is maintained near $\omega_{\rm orb}$ but not exactly at a value equal to $\omega_{\rm orb}$.

  \begin{figure}
   \centering
    \includegraphics[width=8.26cm]{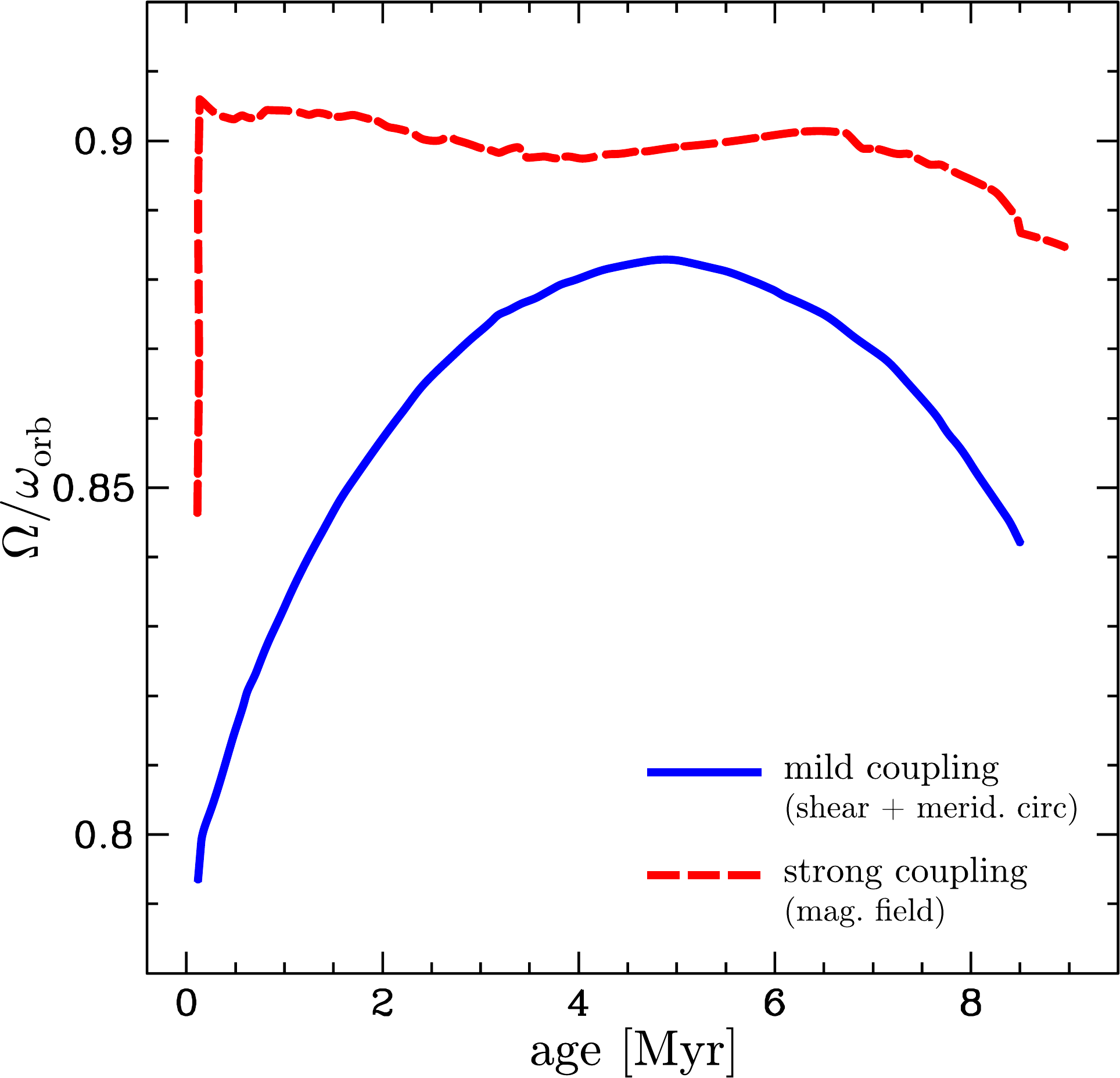} 
      \caption{Evolution as a function of time of the ratio of the surface angular velocity to the orbital velocity for a 15 M$_\odot$ model at Z=0.014 with an initial rotation of
      135 km s$^{-1}$ and with an orbital period of 1.4 days.  The continuous lower curve corresponds to a model without magnetic field and the dashed upper curve corresponds
      to a model with an internal magnetic field.}
         \label{om15su}
   \end{figure}

In Fig.~\ref{dhr15comp}, the evolutionary tracks for the single 15 M$_\odot$ at Z=0.014, with an initial rotation on the ZAMS equal to $\sim$420 km s$^{-1}$ is shown, together
with the tracks of models with the same characteristics but now having a nearby 10 M$_\odot$ companion (orbital period equal to 1.4 days). 
 One case is the same as the one presented in paper I and the other is computed with the Tayler-Spruit dynamo.

When no magnetic field is accounted for, as in paper I, we see that the model in the close binary system is more rapidly mixed than the single star model, hence the small shift to the left with respect to the single star model.
The additional mixing in the binary model is due to the tidally induced shear mixing as explained in  \citet{Song13}.

When the Tayler-Spruit dynamo is accounted for we see that the single star model is more mixed than the corresponding non-magnetic single star model. We see even that the mixing
in that case is so efficient that the star follows the characteristic track of a nearly homogeneous star. This is in line with the results obtained by \citet{MM05}, who indeed noted that
models with magnetic fields were more strongly mixed than models without magnetic field, all other ingredients being kept the same. 
In this model, and this will also be true in the case of the binary model that we have discussed in the present paper, the mixing of the chemical elements is mainly governed by $D_{\rm eff}$, which itself depend on $U$ and thus on $\Omega$. 
This means that in these models, the value of $\Omega$, and not that of the gradient of $\Omega$ \citep[as in models of][]{Song13}, governs the transport of the chemical elements. 

The binary model computed with the Tayler-Spruit dynamo is spined down by the synchronization process. Since $\Omega$ is lower than  in the single star model, the stellar model is less mixed. 
{\it Thus we see that a tidally induced spin-down triggers a more efficient mixing than in a single star  in case the angular momentum is transported by shear and meridional currents and
a less efficient mixing than in a single star when the angular momentum transport imposes a solid body rotation\footnote{Whatever is the physical cause imposing this solid body rotation.}.} 

This is correct for the present 15 M$_\odot$ model which has weak stellar winds.
When stellar winds are strong, then we saw in the present paper that a spin-down can trigger more mixing provided
the angular momentum after synchronization is high enough.
Actually, as we have explained in Section 3, the spin-down in these
systems is rapidly replaced by a spin-up process triggered by strong mass losses.


An interesting consequence of this fact is that these models with strong coupling will never be able to explain
those stars observed to be non-evolved, highly enriched and slowly rotating stars \citep[see group 2 stars discussed in] [] {Hunter09}. Models with milder coupling as those studied in paper I
or with a wind-magnetic braking mechanism  \citep{Meynet11}, in contrast, can produce stars showing these characteristics.


We note that the binary model with magnetic field is more mixed than the binary non-magnetic model.  
This is expected since the magnetic models are more mixed.

On the right panel of Fig.~\ref{dhr15comp}, the evolution of the surface velocity is shown together with some indications on the internal gradient of angular velocity.
The two single star models correspond to the upper lines. 
The magnetic coupling maintains a higher surface velocity than the coupling due only to shear and meridional currents.
The single star model with an internal magnetic field has a very small contrast between the angular velocity at the center and the angular velocity at the surface.
In case of the close binary systems, the surface velocities rapidly converge in the two cases toward more or less the same values as a result of the tidal forces.
The magnetic model, as in the single star case, show very weak contrasts between the rotation of the core and that of the envelope.
   
 \begin{figure*}
   \centering
    \includegraphics[width=8.26cm]{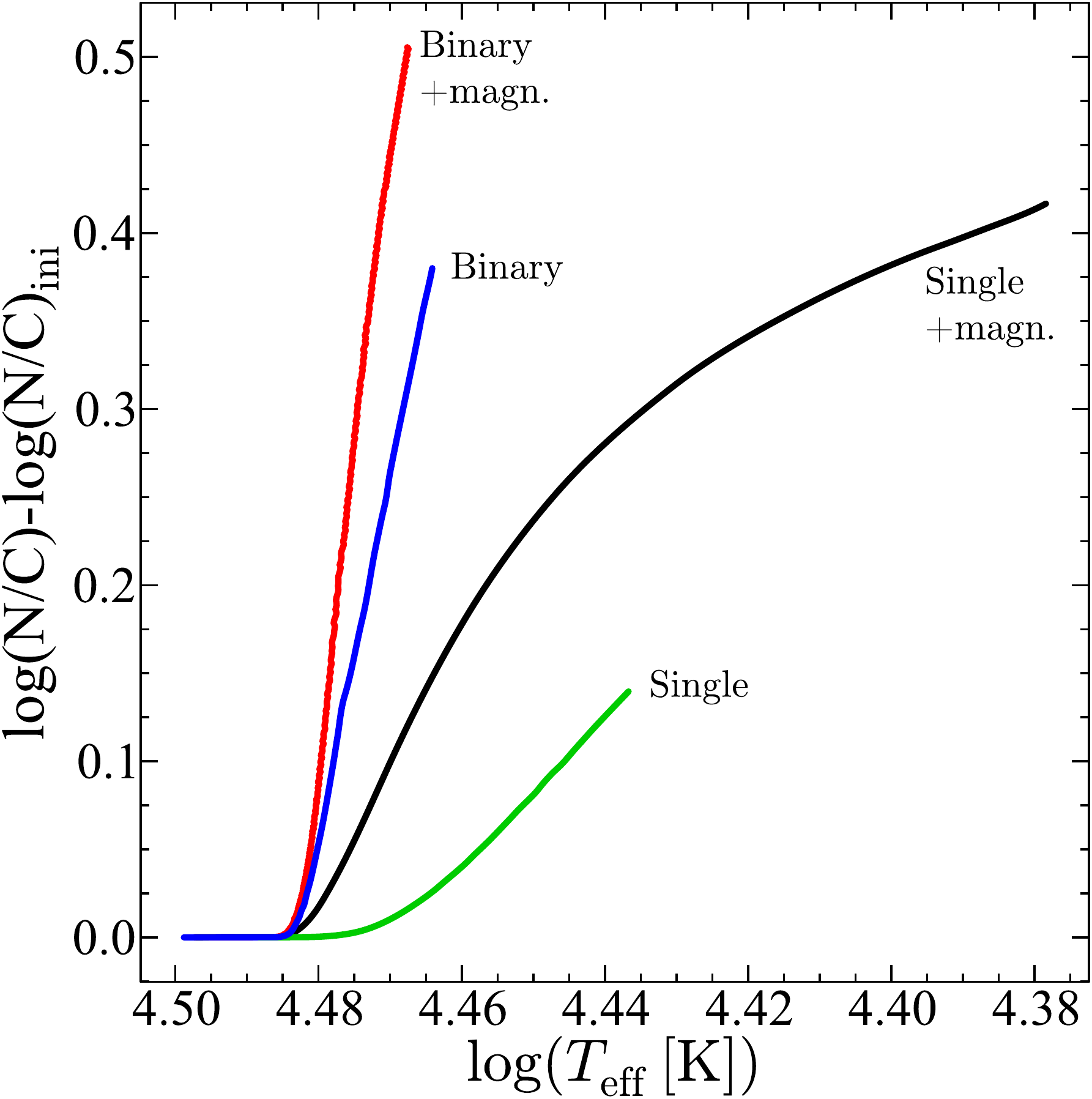}\hfill\hfill  \includegraphics[width=9.3cm]{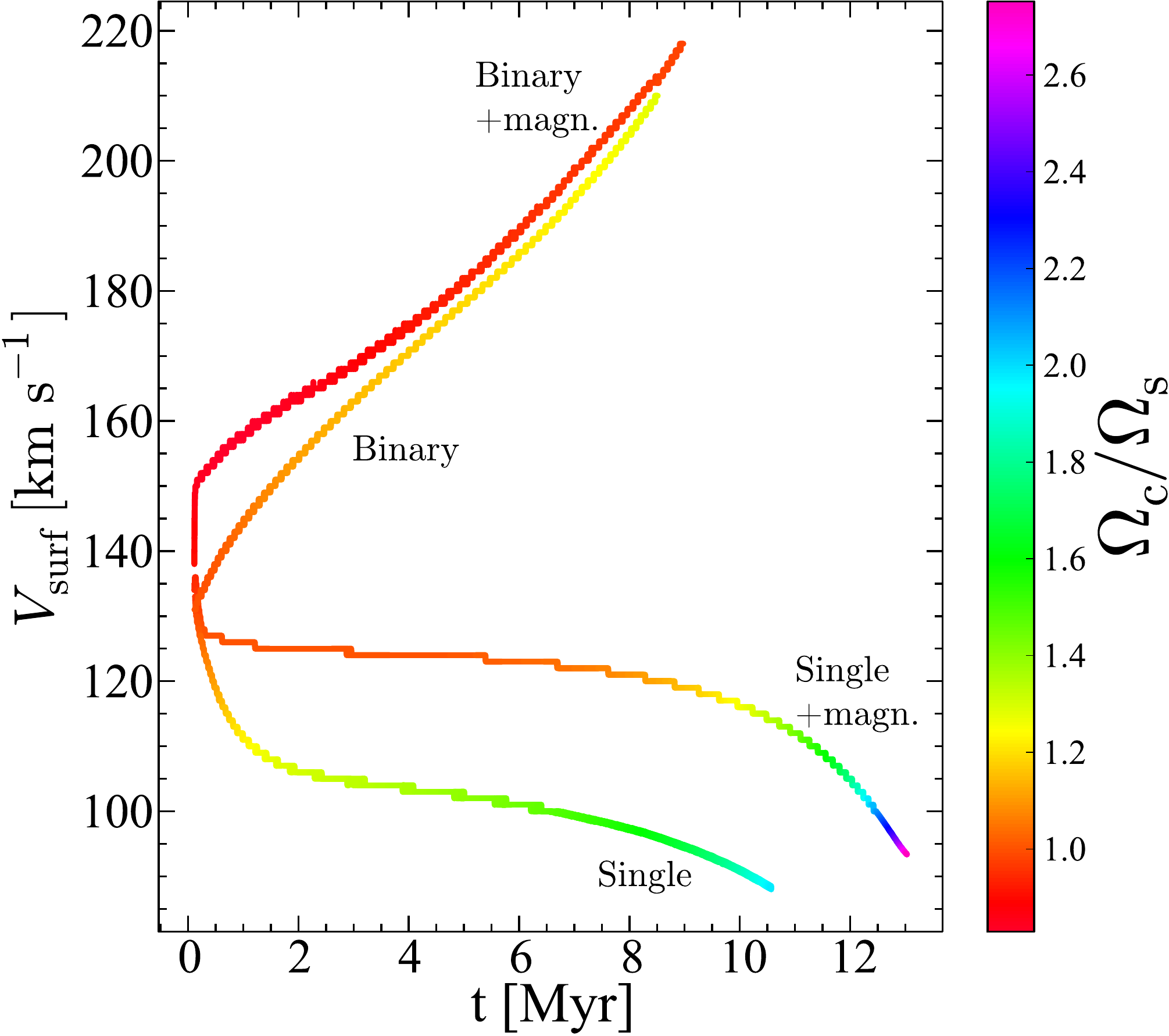}   
      \caption{{\it Left panel:} Evolution of the abundance ratio N/C at the surface of 15 M$_\odot$ models at Z=0.014 with an initial equatorial velocity of $\sim$ 135 km s$^{-1}$ as a function of $\lg T_{\rm eff}$.  Binary models correspond
      to 15 M$_\odot$ with the same characteristics as indicated above but with a 10 M$_\odot$ companion and an initial orbital period of 1.4 days.
      {\it Right panel:} Evolution of the surface equatorial velocity as a function of time for the same models as in the left panel. The colors indicate the ratio between the central
      angular velocity and the surface one.}
         \label{m15su}
   \end{figure*}

\subsection{Spin-up case}  

Let us now compare the cases with and without magnetic field in case of spin-up.
In Fig.~\ref{om15su}, the evolution as a function of time of the ratio of the surface angular velocity to the orbital velocity is shown for the magnetic and the non-magnetic model. We see that
the magnetic model stabilizes very rapidly around a value equal to 90\% the orbital velocity. The non-magnetic model takes more time to reach the maximum value equal to 88\% the orbital period before
to evolve away from it. 

Why does the surface angular velocity of the star never reach the orbital one? As noted above, the evolution of the surface angular velocity results from many effects:
tidal forces tend to make $\Omega$ equal to $\omega_{\rm orb}$, accelerating the star. On the other hand, the inflation of the radius,
the transport of angular momentum inwards due to meridional currents and the loss of angular momentum by stellar winds slow down the surface. 
The evolution shown in Fig.~\ref{om15su} result from all these processes. The fact that the surface angular velocity of the  star only approaches the orbital one indicates that the acceleration coming from the tidal interactions cannot compensate in these models for the counteracting effects of the other processes, thus the synchronization can only be achieved up to a certain point. We note however that observation could hardly detect very precisely  this kind of situation since when the surface angular velocity is already equal to 85-90\% of the orbital one, it will be difficult to say that
it is not equal to the orbital velocity in view of the uncertainties pertaining the measures of the surface rotations of stars and of their radii (in order to obtain the surface angular velocity).
The decrease of $\Omega$ towards the end of the computation results in part from larger mass loss rates and from more rapid inflation of the star.

In Fig.~\ref{om15su}, we can see also that the acceleration of the surface is more rapid in the magnetic than in the non-magnetic model.
 For the spin-down case, we just saw above, that the synchronization timescale increases
when the coupling is stronger. Here we have the inverse situation. Why?
In models with no magnetic field, we have that the meridional currents transport
angular momentum from the surface to the core, hiding thus important amount of angular momentum deposited at the surface in the central region.
This tends to slow down the synchronization process. In solid body rotating models, the core cannot rotate faster than the envelope, preventing
the angular momentum to accumulate in the core. In that case, synchronization is more rapid.
 {\it Thus we see that in case of spin-down, a strong coupling makes the synchronization time longer than in cases of milder coupling, while in case of spin-up, a strong coupling makes the synchronization time shorter.}

The evolutionary tracks in the HRD for the 15 M$_\odot$ stellar models at Z=0.014 with an initial rotation rate around 135 km s$^{-1}$ are very similar
for isolate stars with or without an internal magnetic field. The same is true for the tracks of these same models in a close binary system
with a 10 M$_\odot$ companion and having an initial orbital period of 1.4 days. This is why we do not show these tracks in a figure.
The tracks do not differ much because these systems keep relatively low velocities even in the case of spin-up by tidal forces. These models
follow classical evolution and not a homogeneous evolution.

 On the other hands, the models present at a given age very different values at the surface for the N/C ratio and the equatorial velocity as can be seen in Fig.~\ref{m15su}.
We see that magnetic models are more mixed than non-magnetic ones (whatever single or in a binary system), that binary models are more mixed than single ones (whatever magnetic or not)
and that the magnetic binary model is more mixed than the non-magnetic binary model.
This last point is in contrast with the case of spin-down where the magnetic binary model was less mixed than the non-magnetic binary model. 
How can we explain this difference?
As seen above, the spin-up timescale in the magnetic model is shorter than in the non magnetic one, thus
the magnetic models reach earlier a level of higher rotation and thus has a higher mixing efficiency.

\bibliographystyle{aa}
\bibliography{Song-2}

\end{document}